\documentclass[a4paper,fleqn,usenatbib,useAMS]{mnras}

\usepackage[T1]{fontenc}
\usepackage{ae,aecompl}

\usepackage{graphicx}	% Including figure files
\usepackage{amsmath}	% Advanced maths commands
\usepackage{amssymb}	% Extra maths symbols
\usepackage{hyperref}
\title[Timing analysis of W2R 1926+42]{Kepler light curve analysis of the blazar W2R 1926+42}
\author[Mohan et al.]
{P. Mohan$^{1,2}$\thanks{E-mail: pmohan@shao.ac.cn}, 
Alok C.\ Gupta$^{2}$,
Rumen Bachev$^{3}$,
Anton Strigachev$^{3}$ \\ \\
$^{1}$Shanghai Astronomical Observatory (SHAO), 80 Nandan Road, Shanghai 200030, Shanghai, China.\\
$^{2}$Aryabhatta Research Institute of Observational Sciences (ARIES), Manora Peak, Nainital 263002, India.\\
$^{3}$Institute of Astronomy with NAO, Bulgarian Academy of Sciences, Sofia 1784, Bulgaria.
}

\date{Accepted 2015 November 16.  Received 2015 October 27; in original form 2015 August 19}

\pubyear{2015}

\begin{document}
\label{firstpage}
\pagerange{\pageref{firstpage}--\pageref{lastpage}}
\maketitle

\begin{abstract}
We study the long term Kepler light curve of the blazar W2R 1926+42 ($\sim$ 1.6 years) which indicates a variety of variability properties during different intervals of observation. The normalized excess variance, $F_{\rm var}$ ranges from 1.8 \% in the quiescent phase and 43.3 \% in the outburst phase. We find no significant deviation from linearity in the $F_{\rm var}$-flux relation. Time series analysis is conducted using the Fourier power spectrum and the wavelet analysis methods to study the power spectral density (PSD) shape, infer characteristic timescales and statistically significant quasi-periodic oscillations (QPOs). A bending power law with an associated timescale of $T_B = 6.2^{+6.4}_{-3.1}$ hours is inferred in the PSD analysis. We obtain a black hole mass of $M_\bullet = (1.5 - 5.9) \times 10^7 M_\odot$ for the first time using $F_{\rm var}$ and the bend timescale for this source. From a mean outburst lifetime of  days, we infer a distance from the jet base $r \leq 1.75$ pc indicating that the outburst originates due to a shock. A possible QPO peaked at 9.1 days and lasting 3.4 cycles is inferred from the wavelet analysis. Assuming that the QPO is a true feature, $r = (152 - 378)~ G M_\bullet/c^2$ and supported by the other timing analysis products such as a weighted mean PSD slope of $-1.5 \pm 0.2$ from the PSD analysis, we argue that the observed variability and the weak and short duration QPO could be due to jet based processes including orbital features in a relativistic helical jet and others such as shocks and turbulence.
\end{abstract}


\section{Introduction}

Blazars are a class of radio loud active galactic nuclei (AGN) characterized by strong multi-wavelength timing variability in the complete EM spectrum. 
%and often featureless spectra dominated by continuum emission. 
Blazars are classified into BL Lacertae objects whose spectrum indicates absent or very weak emission lines with typical equivalent widths of $<$ 5 \AA~\citep{1991ApJS...76..813S,1996MNRAS.281..425M}, and flat spectrum radio quasars which show prominent emission lines. 
%both of which display a strong core dominated emission. 
The 
%BL Lacertae objects 
Blazars can be sub-divided into the high-frequency BL Lacs (HBLs) where the synchrotron emission peaks in the UV/X-rays and the low-frequency BL Lacs (LBLs) where the synchrotron emission peaks in radio to optical frequencies. Blazar light curves often indicate rapid, aperiodic variability (e.g. \citealt{1993ARA&A..31..717M,1995ARA&A..33..163W}) or quasi-periodic variability over a diverse range of timescales: $\sim$ 100 s to a few 100 s in the $\gamma$-rays (e.g. \citealt{2007ApJ...664L..71A}); $\sim$ 1000 s to a few hours in the optical/UV and X-rays (e.g. \citealt{2003ApJ...596..847B,2009ApJ...690..216G,2009A&A...506L..17L,2012NewA...17....8G}); intra-day variability in the optical (e.g. \citealt{1996A&A...305...42H,2008AJ....135.1384G,2012MNRAS.425.1357G,2012MNRAS.425.3002G} and references therein) and radio (e.g. \citealt{2013A&A...555A.134L}), short timescale variability of a few days to weeks in optical (e.g. \citealt{2006A&A...450...39G,2012AJ....143...23G} and references therein) and months to years in the optical and radio (e.g. \citealt{2003ApJ...599L..65P,2015MNRAS.452.2004M}). Their spectral energy distribution is mainly composed of non thermal synchrotron and inverse Compton based emission. When interpreted in terms of AGN unification models \citep{1993ARA&A..31..473A,1995PASP..107..803U}, their emission is relativistically beamed due to the following effects: the emitting region is moving towards the observer line of sight at relativistic velocities; the angle towards the observer line of sight is very small (typically $<$ 1$^\circ$); and the jet opening angle (typically $<$ 10$^\circ$, e.g. \citealt{2009A&A...507L..33P}) is very small leading to a highly collimated jet. 

The serendipitous discovery and continuous multi-wavelength monitoring of known blazars has been an ongoing activity in the past few decades with the advent of radio and optical observations of these highly variable sources. Motivations for multi-band optical photometric monitoring include understanding the causes for variability over diverse timescales; inferring jet properties including the Doppler factor, magnetic field strength, kinematics of components, the shape of the spectrum (e.g. \citealt{2015MNRAS.450..541A,2015MNRAS.451.3882A} and references therein) in aiding spectroscopic studies as well as determining the emitting region size and distance from the central engine; aiding multi-wavelength campaigns which study correlated variability during flaring events (e.g. \citealt{2012MNRAS.425.1357G,2013A&A...558A..92B,2015ApJ...807...79H,2015A&A...576A.126A} and references therein) and in construction of the spectral energy distribution of the blazar (e.g. \citealt{2009A&A...507..769R}) to comment on emission mechanisms; in the search for quasi-periodic oscillations (e.g. \citealt{2009ApJ...690..216G}) and in the study of the power spectral density shape, both of which can be used to constrain theoretical models and derive jet parameters (e.g. \citealt{2014JApA...35..431M,2014ApJ...791...21F,2015ApJ...805...91M}). Optical monitoring of blazars is often focused on studies of intra-day variability in prominent individual sources such as S5 0716+714, 3C 454.3, PKS 2155-304, Mrk 421 and others (e.g. \citealt{2004MNRAS.350..175S,2008AJ....135.1384G,2009AJ....138.1428F,2012MNRAS.425.3002G,2014MNRAS.443.2940H}) and the description of the optical spectrum shape from the integrated flux to achieve the above goals.

The identification and classification of AGN in the Kepler field was based on surveys conducted by the Wide-field Infrared Survey Explorer (WISE), Two Micron All-Sky Survey (2MASS), and ROSAT all-sky survey (RASS) \citep{2011ApJ...743L..12M,2012ApJ...751...52E}. These include both radio quiet and radio loud sources. Follow up studies include the determination of black hole mass ($M_\bullet$) using reverberation mapping in the narrow line Seyfert 1, KA 1858+4850 where $M_\bullet = 8 \times 10^6 M_\odot$ ($M_\odot$ is the solar mass) was obtained \citep{2014ApJ...795...38P}; a timing analysis of the Kepler light curves from three radio loud AGN and one Seyfert type 1.5 galaxy where a power spectral density (PSD) slope ranging between -1.6 and -2.0 was obtained with no characteristic timescales in the red noise dominated portion \citep{2014ApJ...785...60R}; and, the application of the continuous-time autoregressive moving average (CARMA) process to model the PSD shape from the Kepler light of the AGN Zw 229-15 indicating an approximate power law PSD shape with slope $\sim$ -3 for timescales $<$ 1 month, consistent with the results of \cite{2011ApJ...743L..12M} and a flattening of the PSD with slope of $-2$ for timescales $>$ 10 days \citep{2014ApJ...788...33K}.

W2R 1926+42 ($\alpha_{2000.0}$ = 19h 26m 31.09s, $\delta_{2000.0}$ = +42$^\circ$ 09$^{'}$ 59.0$^{''}$) was first identified in the Kepler field and classified as a BL Lacertae object by \cite{2012ApJ...751...52E}. In their study, the WISE, 2MASS, and RASS data were made use of to identify bright AGN. W2R 1926+42 was identified as a BL Lacertae object with a redshift $z = 0.155$ based on the FEI5269 and NaD5892 absorption lines \citep{2013ApJ...766...16E}. In the same study, a compilation of the radio to X-ray non-simultaneous spectral fluxes based on archival data is used to produce its spectral energy distribution (SED), using which, W2R 1926+42 is classified as an LBL based on the location of its synchrotron peak. It is the only identified BL Lac object in the Kepler field. Indications of fractal behaviour in some blazar light curves motivates a search for low dimensional chaos signatures in the long term Kepler light curve of W2R 1926+42 \citep{2015A&A...576A..17B}, yielding a negative result, which places constraints on the particle acceleration mechanism. Deterministic injection of particles into the jet at the base from the disk-jet region is found to be unlikely to cause the variability in that study. It is then suggested that processes such as magnetic reconnection and turbulence in independently emitting active zones along the jet could be responsible for the observed variability.

\section{Data extraction and preliminary properties}

\begin{figure}
\centerline{\includegraphics[scale=0.45]{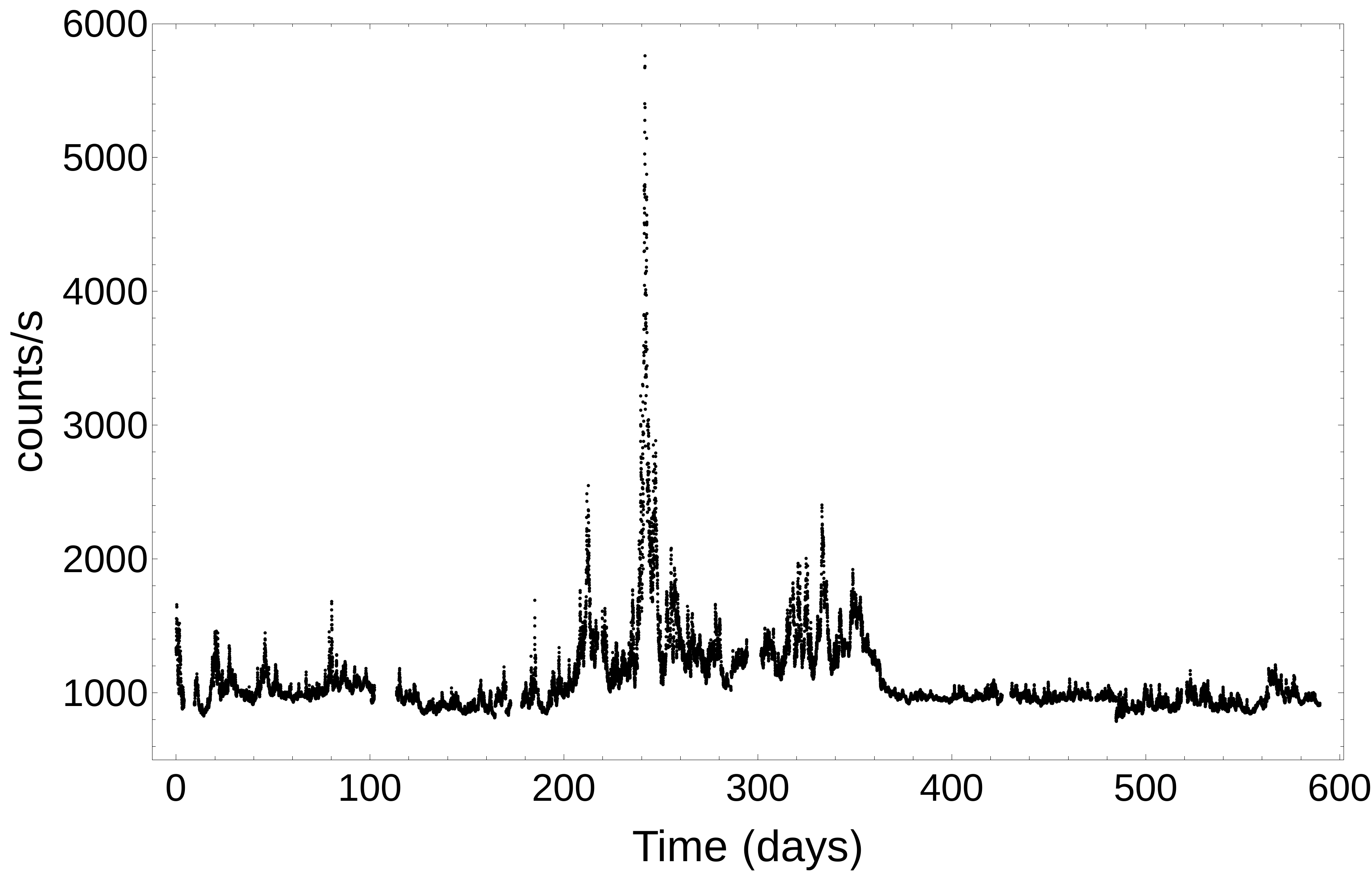}}
\caption{Complete updated Kepler light curve of W2R 1926+42 spanning 589 days ($\sim$ 1.6 years) indicating the quiescent portions from 0-180 days and from 380 days to the end of the observation and strongly flaring portions between 180-380 days.}
\label{fullLC}
\end{figure}

The Kepler light curve of the BL Lac W2R 1926+42 studied by \cite{2013ApJ...766...16E} is of $\sim$ 180 days duration spanning the 11th and 12th quarters of the Kepler observations during which the source was in a relatively quiescent phase. The updated light curve spans 589 days ($\sim$ 1.6 yrs) with a median sampling rate of 0.02 days, containing small data gaps. The entire light curve without binning and interpolation is presented in Fig. \ref{fullLC}. The updated light curve spans the 11th to 17th quarter of the Kepler observations, starting on 29th September 2011 and consists of 18 uniformly sampled segments of length between 1300 and 16000 points covering roughly the wavelength range between 4300 \AA~ and 8900 \AA. A detailed description of the extraction of the light curve including initial pre-processing is presented in \cite{2015A&A...576A..17B}. The light curve is extracted using the standard Kepler procedure (PDSCAP) which is automatically applied to every Kepler source and is available on the \texttt{Kepler data archives} \footnote{http://archive.stsci.edu/kepler/}. Photometric errors are typically $\sim $ 0.2 \%. After the 11th and 12th quarters, there was strong flaring activity lasting between $\sim$ 200 days and 380 days followed by another relatively quiescent phase till the end of the observations. This presence of quiescent and strongly flaring portions including the switch in between with short and long term trends indicates a feature rich light curve, thus, warranting a comprehensive timing analysis. 

\section{Time series analysis}

\subsection{Light curve preparation and variability measurement}

\begin{figure}
\centerline{\includegraphics[scale=0.3]{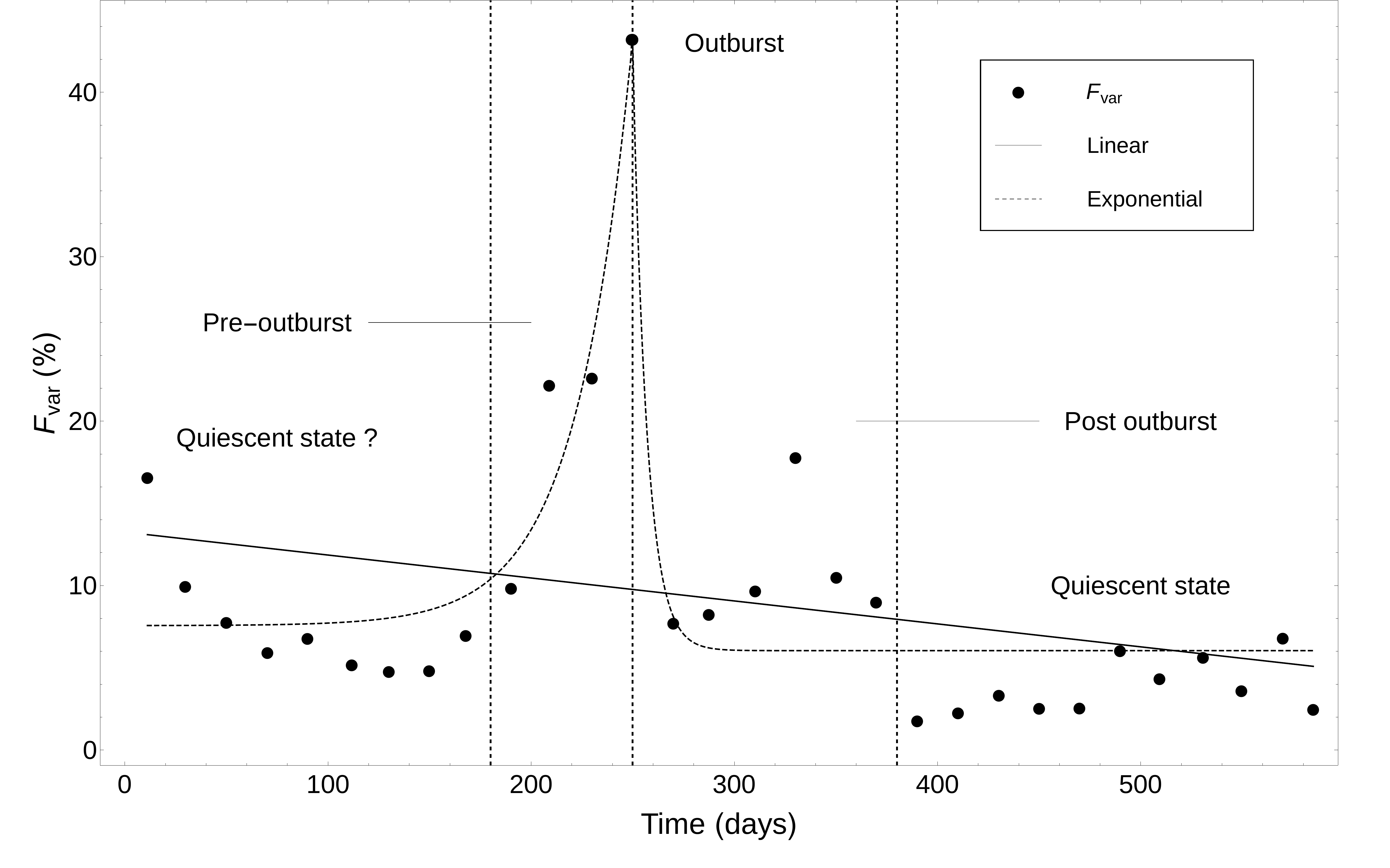}}
\caption{$F_{\rm var}$ as a function of observation time. The measured values follow the variability pattern of the light curve in Fig. \ref{fullLC} and indicate the following distinct regions: a possible quiescent phase (0 - 180 days), a pre-outburst phase (180 - 250 days), an outburst at 250 days, a post-outburst phase (250 - 380 days) containing a smaller outburst at $\sim$ 330 days, followed by a quiescent phase  (380 - 589 days) with $F_{\rm var}$ ranging between 1.8 \% and 43.3 \%. For specific details of $F_{\rm var}$ ranges during each phase and the fits carried out, refer to text.}
\label{Fvart}
\end{figure}

A measure of intrinsic source based variability in the light curve is the normalized excess variance $F_{\rm var}$ (e.g. \citealt{1996ApJ...470..364E,1997ApJ...476...70N,2001ApJ...554..274E,2002ApJ...568..610E}). For a light curve $x(t_n)$ of length $N$ with a mean $\mu$, standard deviation $\sigma$ and measurement error at each point $\sigma_{N} (t_n)$, $F_{\rm var}$ is given by
\begin{equation}
F_{\rm var} = \frac{1}{\mu} \sqrt{\sum^{N}_{n=1} \left[(x(t_n)-\mu)^2-\sigma^2_{N} (t_n)\right]}
\end{equation}
An initial filtering of the light curve to remove small variable features reveals on inspection, typical variability on timescales of $\sim$ 20 days. The original unevenly sampled light curve is thus split into 20 day segments, each with a median time sampling rate of 0.02 days, ensuring that there were sufficient number of points (typically $>$ 300) to measure $F_{\rm var}$ and study its evolution across the observation duration. The measured normalized excess variance $F_{\rm var}$ follows the variability trend in the light curve and is presented in Fig. \ref{Fvart}. 

\begin{figure}
\centerline{\includegraphics[scale=0.15]{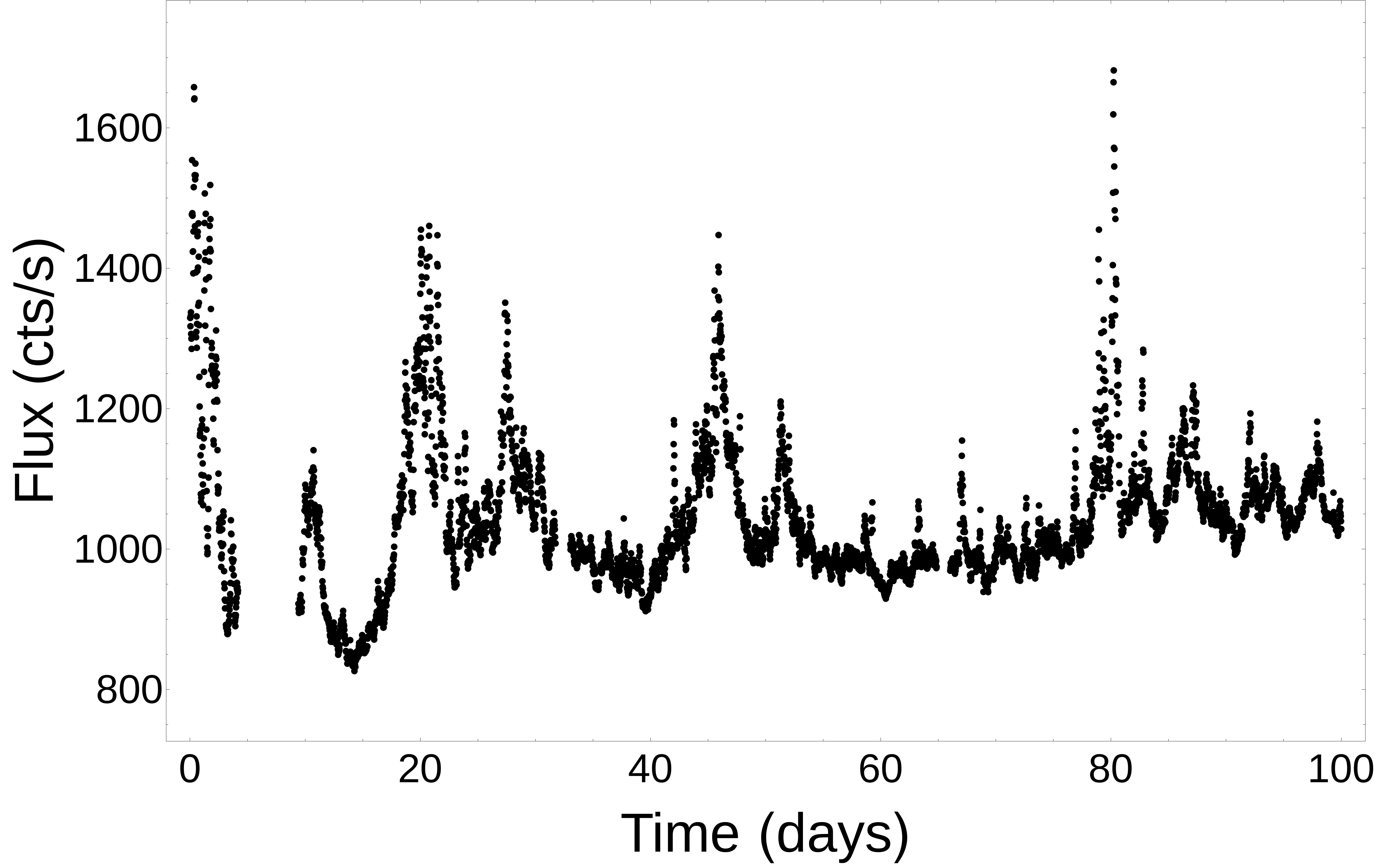}\includegraphics[scale=0.15]{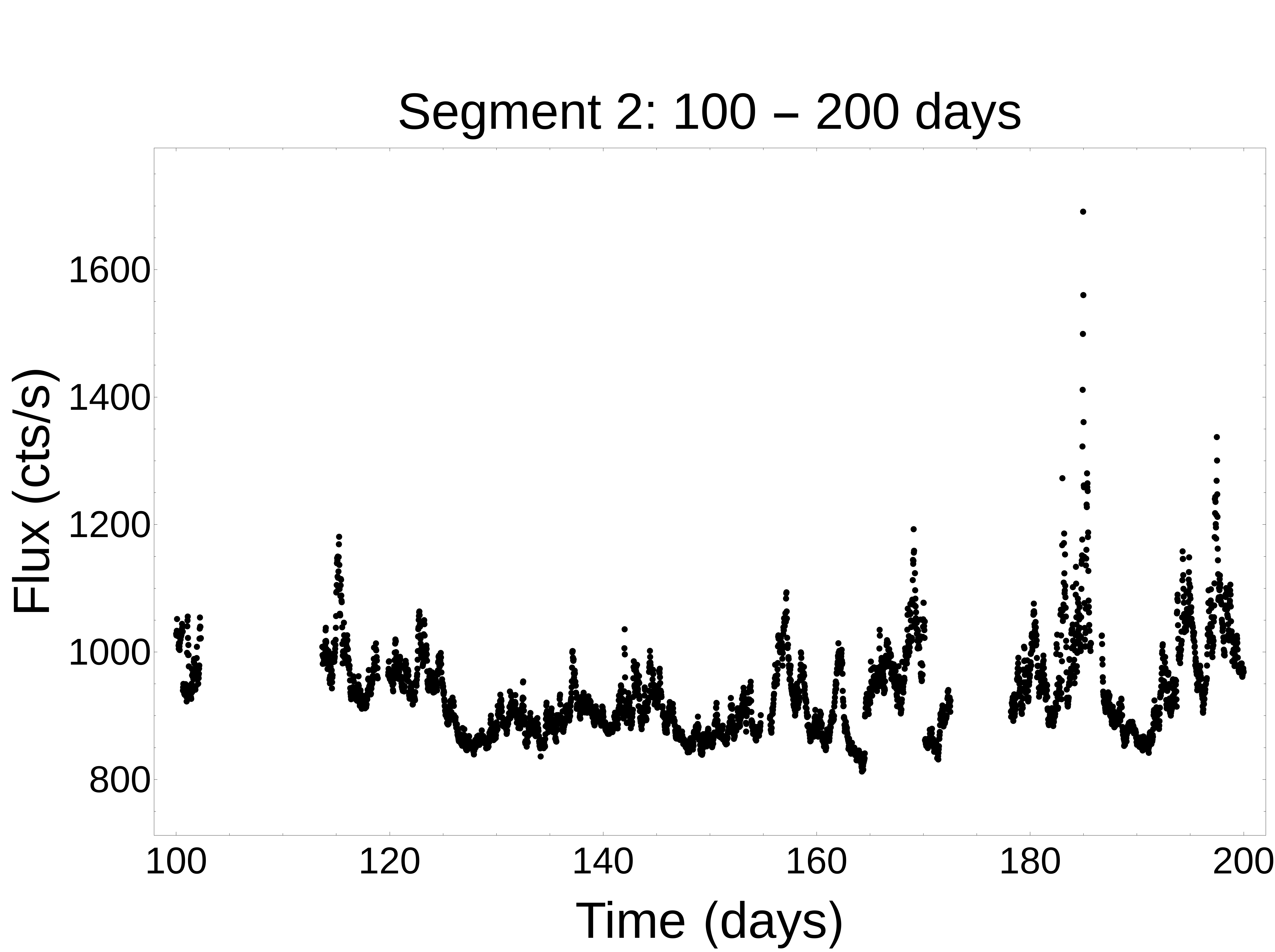}}
\centerline{\includegraphics[scale=0.15]{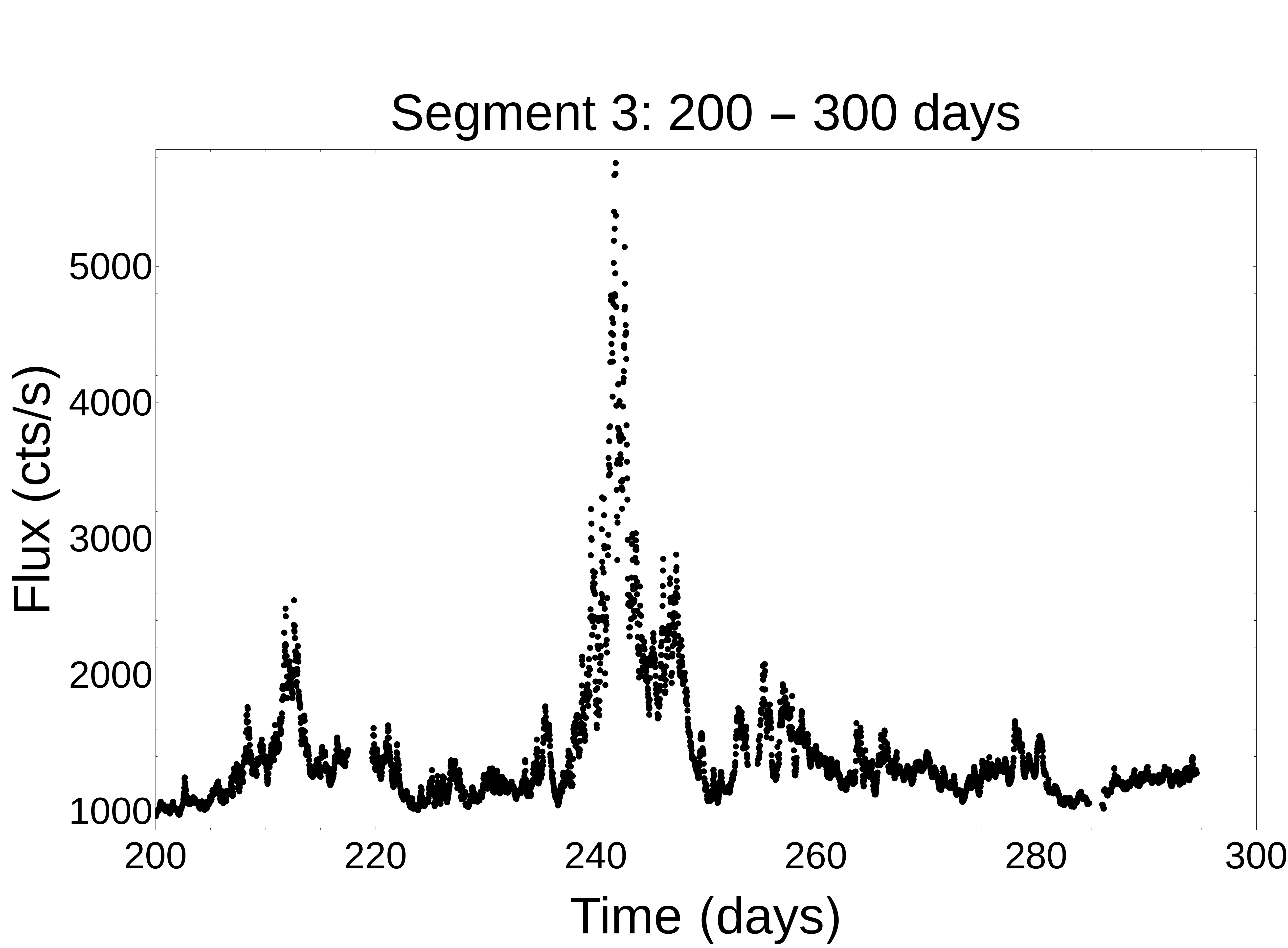}\includegraphics[scale=0.15]{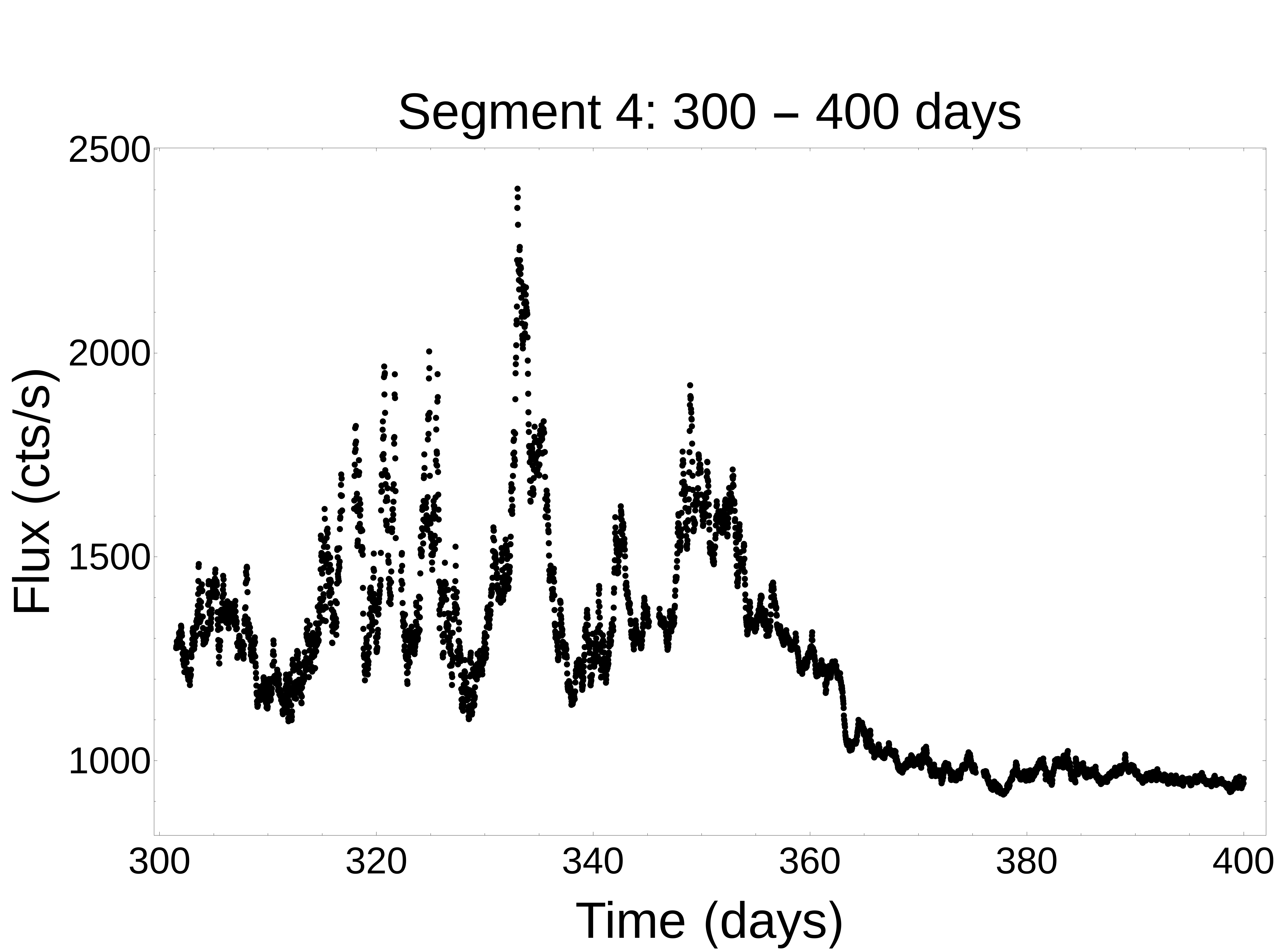}}
\centerline{\includegraphics[scale=0.15]{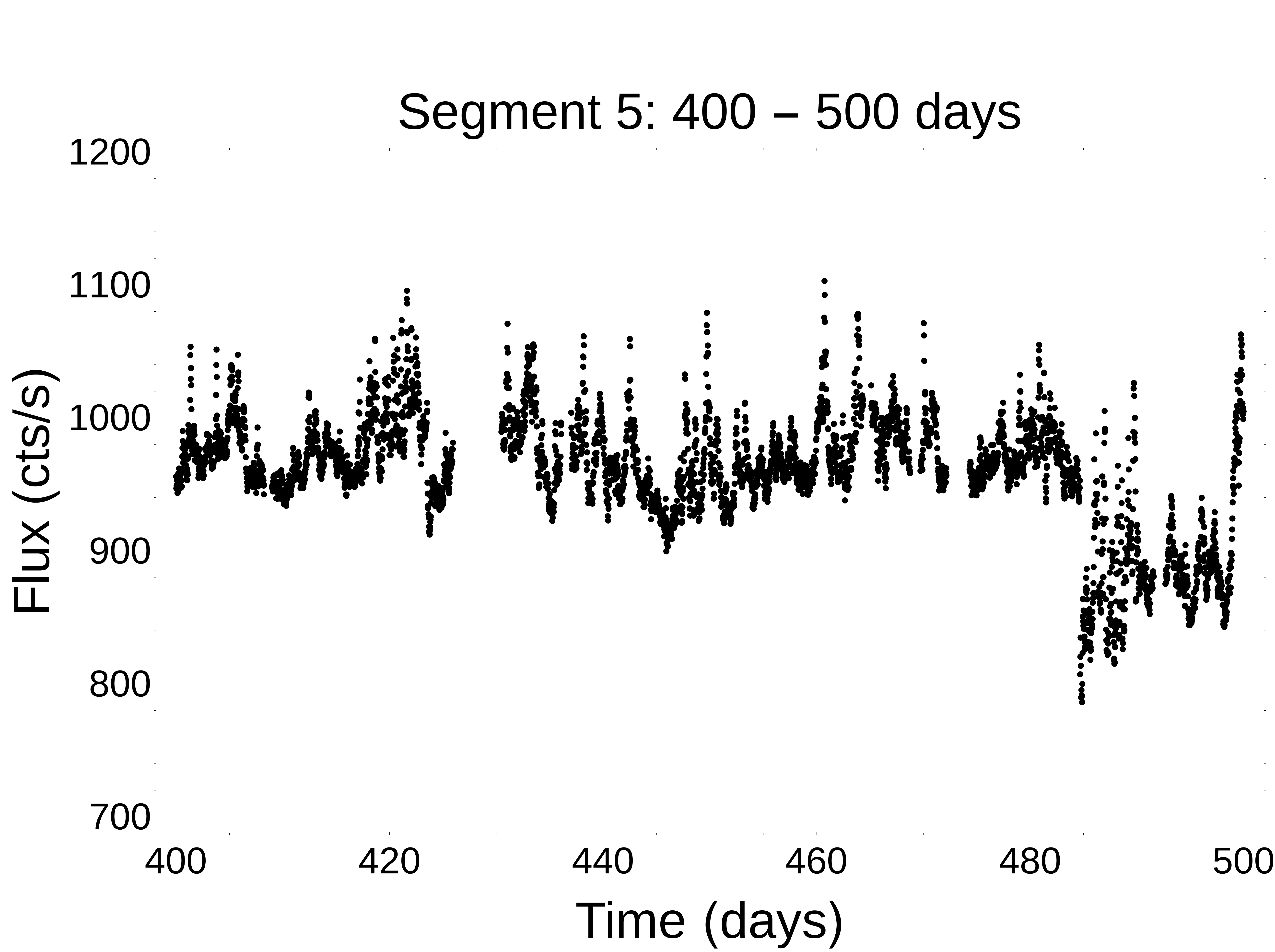}\includegraphics[scale=0.15]{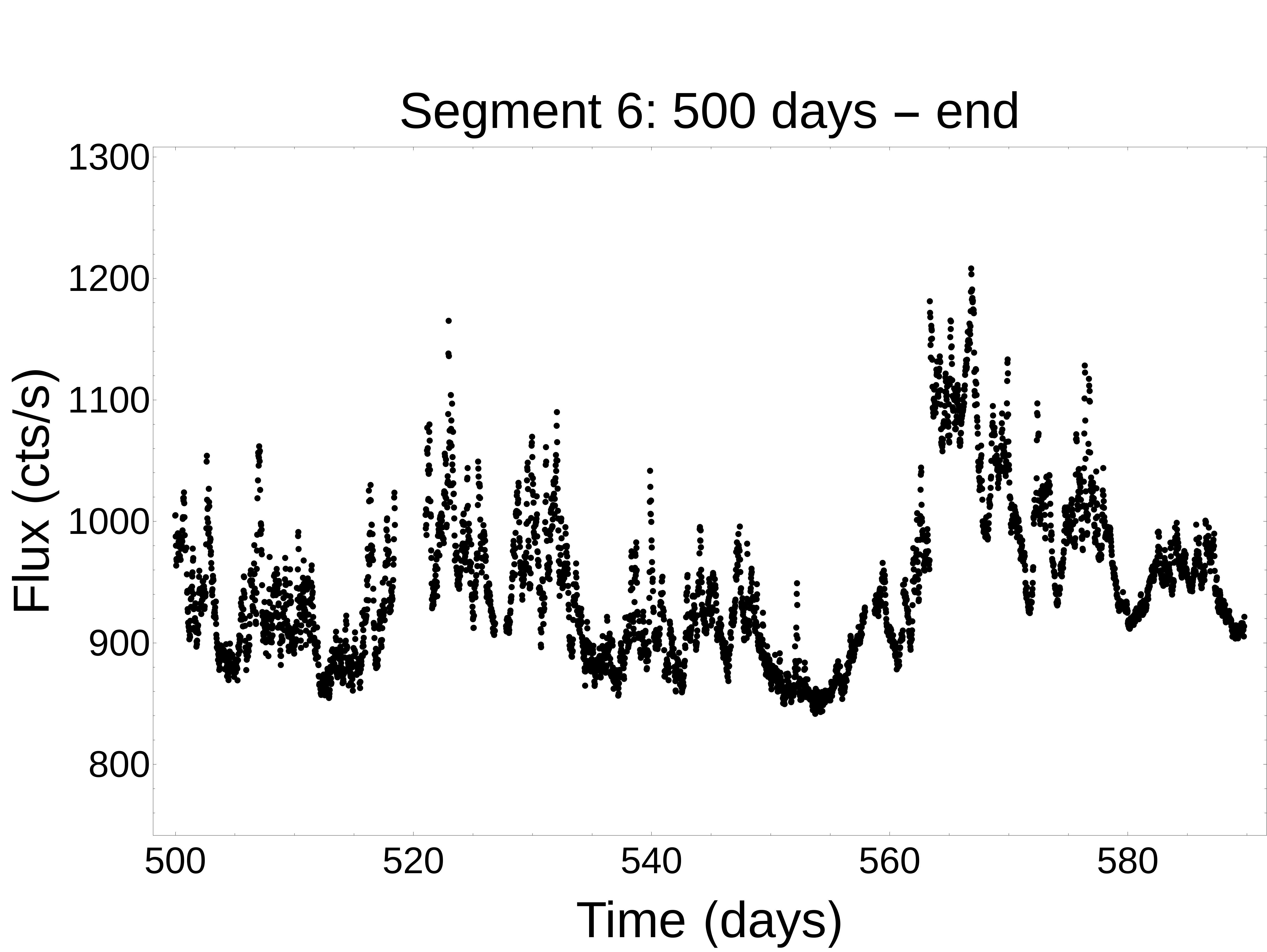}}
\caption{Individual light curves of the segments 1-6 indicating the quiescent and flaring portions as seen by the changes in flux from segment to segment.}
\label{lcplot1}
\end{figure}

The original unevenly sampled light curve is then binned, interpolated and made evenly sampled at regular intervals of 0.2 days for the timing analysis, which includes the search for quasi-periodic oscillations (QPOs) and other characteristic timescales, the evolution of variability properties with time of observation and using this timing information to infer black hole mass and the region of emission along the jet. To conduct these studies, the light curve is split into six segments each spanning $\sim$ 100 days, presented in Fig. \ref{lcplot1}. Splitting of the light curve into these segments was necessitated as the light curve indicates long duration low flux and high flux states. The timing analysis would then be able to capture important features or trends in the light curve. This was also done in order to decrease the computational load.

\subsection{Timing analysis}

We performed a timing analysis of the full interpolated light curve as well as the six segments using the Fourier periodogram and wavelet analysis. The Fourier periodogram analysis involves the determination of the underlying PSD shape, any characteristic timescales and any coherent, statistically significant QPOs. The normalized periodogram is given by,
\begin{equation}
P(f_j)=\frac{2 \Delta t}{\mu^2 N} |F(f_j)|^2
\end{equation}
where $\Delta t$ is the sampling time step for the evenly sampled light curve $x(t_n) = x(n \Delta t)$, and $|F(f_j)|$ is its discrete Fourier transform evaluated at frequencies $f_j = j/(N \Delta t)$ with  $j = 1, 2,..,(N/2-1)$. We used two competing parametric models to constrain the PSD shape. The power law with constant noise model of $P(f_j)$ is given by
\begin{equation}
I(f_j) = A f^{\alpha}_j+C,
\end{equation}
with amplitude $A$, slope $\alpha$, and a constant Poisson noise $C$. The bending power law with constant noise model of $P(f_j)$ is given by
\begin{equation}
I(f_j) = A f^{-1}_j \left(1+(f_j/f_b)^{-\alpha-1}\right)^{-1}+C.
\end{equation}
with amplitude $A$, slope $\alpha$, bend frequency $f_{b}$, and a constant Poisson noise $C$. These two models have been used in literature to model the PSD shape of X-ray and radio light curves (e.g. \citealt{2012A&A...544A..80G,2014ApJ...791...74M,2015MNRAS.452.2004M}). The fit is carried out using the maximum likelihood estimator method and parameters $\theta_k$ for each of the above two models are determined. For a model $I(f_j,\theta_k)$ with parameters $\theta_k$, the minimization of its log-likelihood function (e.g. \citealt{2013MNRAS.433..907E,2014JApA...35..397M,2014ApJ...791...74M}) given by
\begin{equation}
S(\theta_k) = - 2 \sum^{n-1}_{j=1}(\ln(I(f_j,\theta_k))+P(f_j)/I(f_j,\theta_k)),
\end{equation}
is equivalent to the maximization of likelihood and yields the best fit parameters. Once the best fit parameters are determined for each model, model selection is carried out using the Akaike information criteria (AIC) \citep{2014ApJ...791...74M,2015MNRAS.452.2004M} to determine the best fit PSD shape. The AIC is a measure of loss of information when data is fit by a given model. The more the loss, the higher is the entropy and hence, the AIC value. The AIC cannot be used as an absolute test for a model (goodness of fit), but can be used to compare the relative effectiveness of a given model over another in describing the PSD shape adequately. The AIC and model likelihood are defined by
\begin{align}
AIC&=S(\theta_k)+2 p_k, \\ \nonumber
\Delta_i&=AIC_{\mathrm{min (model \ i)}}-AIC_{\mathrm{min(null)}}, \\ \nonumber
L(\mathrm{model \ i|data})&=e^{-\Delta_i/2},
\end{align}
where $p_k$ is the number of $\theta_k$ used in the model; the null model is the simplest model which here is the power law PSD model, and $L(\mathrm{model \ i|data})$ is the likelihood that model $i$ is a better fit compared to the null. Using such a definition, we can study other parametric models in addition to the above two commonly used models. Models with $\Delta_i\leq 2$ can be considered close to the null, those with $4 \leq \Delta_i \leq 7$ are considerably less supported, and those with $\Delta_i > 10$ cannot be supported \citep{2004..Likelihood}. 

For a light curve populated with random Gaussian noise, its PSD ordinates are $\chi^2_2$ distributed (e.g. \citealt{2002apa..book.....F}). The residuals of the fit to the data periodogram are thus expected to be $\chi^2_2$ distributed for the best fit PSD shape and the case of no QPO. We use an analytic significance test based on the $\chi^2_2$ statistics, accounting for the number of frequencies sampled \citep{2005A&A...431..391V} to infer the statistical significance of any detected QPO. As the residual $\gamma(f_j) = P(f_j)/I(f_j)$ is $\chi^2_2$ distributed, the integrated area under the probability density function of the $\chi^2_2$ distribution (gamma density $\Gamma(1,1/2)= \exp{(-x/2)}/2$) upto a limit $(1-\epsilon)$ gives the probability that the power associated with a QPO is different from the rest of the population. When cast in units of the periodogram power, it is given by (\citealt{2005A&A...431..391V}),
\begin{equation}
\gamma_\epsilon=-2 \ln \left[1-(1-\epsilon)^{\frac{1}{N/2-1}}\right],
\end{equation}
accounting for the $N/2-1$ trial frequencies used in the search for the QPO. Once $\epsilon$ is specified, $\gamma_\epsilon$ is calculated and multiplied with $I(f_j)$ to give the significance level used to identify outliers in the periodogram that could indicate the presence of a statistically significant QPO.

The wavelet analysis uses a sampling kernel which can be scaled in size and shifted in the frequency domain (e.g. \citealt{doi:10.1146}) to obtain a power spectrum which is a function of the sampling frequency and time of observation. The wavelet transform of an evenly sampled light curve $x(n \Delta t)$ is given by (e.g. \citealt{1998BAMS...79...61T})
\begin{equation}
W (n,s) = \sum^{N}_{n' = 1} x(n \Delta t) \psi^{\ast} \left(\frac{(n'-n) \Delta t}{s}\right),
\end{equation}
where $\psi^{\ast}$ is the complex conjugate of the wavelet sampling kernel which can be shifted in the time domain using a time parameter $n'$ and can be scaled in size to sample sections of varying length in the light curve using the scaling parameter $s$. Common choices for the wavelet sampling kernel include the Morlet wavelet, derivative of Gaussian wavelet family and others. The Morlet wavelet which we use here is given by $\psi = \pi^{-1/4} e^{i \omega_0 \eta} e^{-\eta^2/2}$ where $\omega_0 = 6$ is a non-dimensional frequency and $\eta$ is a non-dimensional time parameter. The wavelet transform can be written in the frequency domain as the inverse Fourier transform of the convolution product,
\begin{equation}
W (n,s) = \sum^{N}_{j=1} F(\omega_j) \Psi^{\ast} (s \omega_j) e^{i \omega_j n \Delta t},
\end{equation}
where $F(\omega_j)$ is the discrete Fourier transform of $x(n \Delta t)$ evaluated at circular frequencies $\omega_j = 2 \pi j/(N \Delta t)$ with $j = 1, 2,.., N$ and $\Psi^{\ast} (s \omega_j) = \pi^{-1/4} e^{-(s \omega_j-\omega_0)^2/2}$ is the complex conjugate of the Fourier transform of the Morlet wavelet. The wavelet power spectrum is then given by
\begin{equation}
P_W (n,s) = W(n,s).W^{\ast} (n,s),
\end{equation}
where $W^{\ast} (n,s)$ is the complex conjugate of the wavelet transform. Our implementation of the wavelet analysis is based on the algorithm prescribed by \cite{1998BAMS...79...61T} using the Morlet wavelet sampling function. The analysis can be used to detect QPOs and study their evolution, e.g a possible 4.6 hour periodicity lasting 3.8 cycles inferred in the 64 ks XMM-Newton X-ray light curve of the blazar PKS 2155-304 \citep{2009A&A...506L..17L}. 

The best fit PSD shape is used in Monte-Carlo simulations based significance testing to determine the statistical significance of any detected quasi-periodic components in the wavelet analysis, the procedure of which is based on a search and data characterization strategy presented in \cite{2014JApA...35..431M} and implemented in \cite{2012MNRAS.425.1357G} for the wavelet analysis of the optical light curve of the blazar S5 0716+714. The \cite{1995A&A...300..707T} algorithm and the $\chi^2_2$ statistics are used to simulate power spectra with similar statistical and variability properties as the original light curve. The collapsed global power spectrum, which is the weighted sum of all power along a given horizontal time slice (weighted by the total power in that slice) in the wavelet power spectrum corresponds to a smoothed version of the Fourier power spectrum. For a given power peak indicating a QPO in the data, we determine the number of times, $n$ that a peak in simulated global wavelet power at that same position exceed the true power peak value. If the number of simulations is $N$, the reported significance of a particular peak is $(1-n/N)~100$\%. Using this strategy, we can test for statistical significance of detected wavelet peaks and can also use the entire time slice even in the presence of edge effects due to the cyclic nature of the wavelet sampling process.
%The duration of power peaks includes only those in the region within a cone of influence (black triangle shaped region in the wavelet power spectrum. e.g. for segment 1 in Fig. \ref{waveletseg1}) as those outside it could be affected by edge effects due to the cyclic nature of the sampling wavelet process. 

\section{Results}

\subsection{Normalized excess variance}

\begin{figure}
\centerline{\includegraphics[scale=0.3]{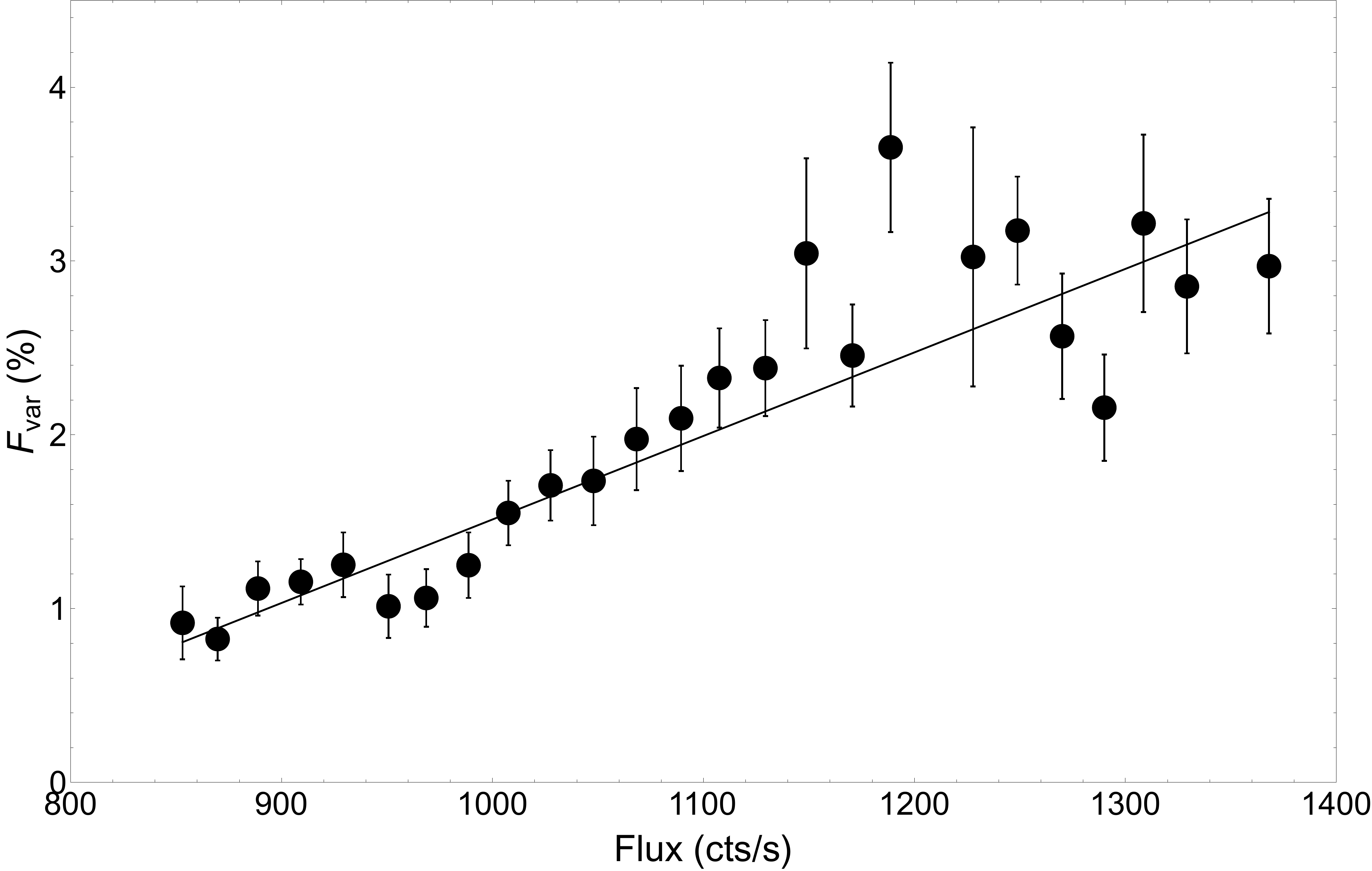}}
\caption{Binned $F_{\rm var}$ as a function of flux. The linear model indicates a reasonable fit with $\chi^2/dof = 1.30$ (29.98/23 dof). No deviation from linear trend is inferred.}
\label{Fvarflux}
\end{figure}

The normalized excess variance $F_{\rm var}$ ranges from a minimum of 1.8 \% in the quiescent phase after $\sim$ 400 days to a maximum of 43.3 \% during the strongly flaring phase between $\sim$ 180 days and 380 days, following the variability trend in the light curve and presented in Fig. \ref{Fvart} indicating a relation between the variance and the flux.
%, consistent with the linear $F_{\rm var}$-flux relation inferred in \cite{2013ApJ...766...16E} for this source. 
To study the $F_{\rm var}$-flux relation, we first divided the light curve into 2360 segments, each with an average time duration of 0.25 days. We discarded segments where there were less than 10 points which reduced the number of segments to 2105. The mean and $F_{\rm var}$ was then calculated for each of these segments. As the data in this form was noisy, we binned the $F_{\rm var}$ in flux bins, ensuring that each bin contained $>$ 20 points. We obtained 25 bins and the $F_{\rm var}$-flux points were fit with a linear model which indicates a reasonable fit with $\chi^2/dof = 1.30$ (29.98/23). The $F_{\rm var}$-flux relation inferred from our study is presented in Fig. \ref{Fvarflux}. A possible deviation from a linear relation was inferred in \cite{2013ApJ...766...16E}. In our analysis, we infer a consistency with linearity considering that our light curve is longer and hence, there are more points to populate the parameter space. We do see the same deviation at $\sim$ 1200 cts/s, though, this appears to be a deviation from the mean trend. The $F_{\rm var}$ measured for the entire light curve is 27.8 \%. Major variability phases in the light curve are studied using the measured $F_{\rm var}$ as a function of observation time. The variability trend in the light curve on inspection can be classified into a possible quiescent phase (0 - 180 days) with $F_{\rm var}$ range of 4.8 \% to 16.6 \%, a pre-outburst phase (180 - 250 days) with $F_{\rm var}$ range of 9.9 \% to 22.7 \%, an outburst at 250 days with $F_{\rm var} = $43.3 \%, a post-outburst phase (250 - 380 days) containing a smaller outburst at $\sim$ 330 days with $F_{\rm var}$ range of 7.7 \% to 17.8 \%, followed by a quiescent phase  (380 - 589 days) with $F_{\rm var}$ range of 1.8 \% to 6.8 \%. The $F_{\rm var}$ as a function of observation time indicates a decreasing trend on inspection. When the data points are fit with a linear model of the form $F_{\rm var} = m~t_i+c$ and obtain a slope $m = -0.014 \pm 0.009$ day$^{-1}$ and a normalization $c = 13.25 \pm 3.01$. The $F_{\rm var}$ between 0 - 250 days is fit with an exponential model of the form $F_{\rm var} = A~ e^{(t_i-t_0)/\tau_R}+B$ where $t_0$ is the time of outburst, $\tau_R$ is the outburst rise time, $A$ is the exponential normalization and $B$ is a linear normalization for this section. We obtain $A = 35.61 \pm 4.27$, $B = 7.56 \pm 1.57$ and $\tau_R = 27.6 \pm 7.4$ days. The $F_{\rm var}$ between 250 - 589 days is fit with an exponential model of the form $F_{\rm var} = A ~e^{(t_0-t_i)/\tau_D}+B$ where $\tau_D$ is the outburst decay time, $A$ is the exponential normalization and $B$ is a linear normalization for this section. We obtain $A = 37.20 \pm 4.38$, $B = 6.04 \pm 1.07$ and $\tau_D = 7.0 \pm 5.1$ days. With weights $w_R= 1/\sigma^2_R = 1/7.4^2$ day$^{-2}$ and $w_D = 1/\sigma^2_D = 1/5.1^2$ day$^{-2}$, the weighted mean flare lifetime is 
\begin{equation}
\tau = \left(\frac{w_R \tau_R+w_D \tau_D}{w_R+w_D}\right) {\rm day},
\end{equation}
and the error estimate is 
\begin{equation}
\sigma_\tau = \left(\frac{w_R (\tau_R-\tau)^2+w_D (\tau-\tau_D)^2}{w_R+w_D}\right)^{1/2} {\rm day}.
\end{equation}
We obtain a mean weighted flare lifetime of $\tau = 13.6 \pm 9.6$ days. The measured $\tau_R > \tau_D$ indicates an asymmetric shape of this prominent flare with a longer rise time compared to a short decay time, though when considering the measure of asymmetry on a whole, the light curve is consistent with being roughly symmetric \citep{2015ApJ...805...80C}. The $F_{\rm var}$ ranges, the relevant phase and any characteristic timescales inferred from our study are summarized in Table \ref{Fvar}.

\subsection{Fourier periodogram analysis}

\begin{figure}
\centerline{\includegraphics[scale=0.2]{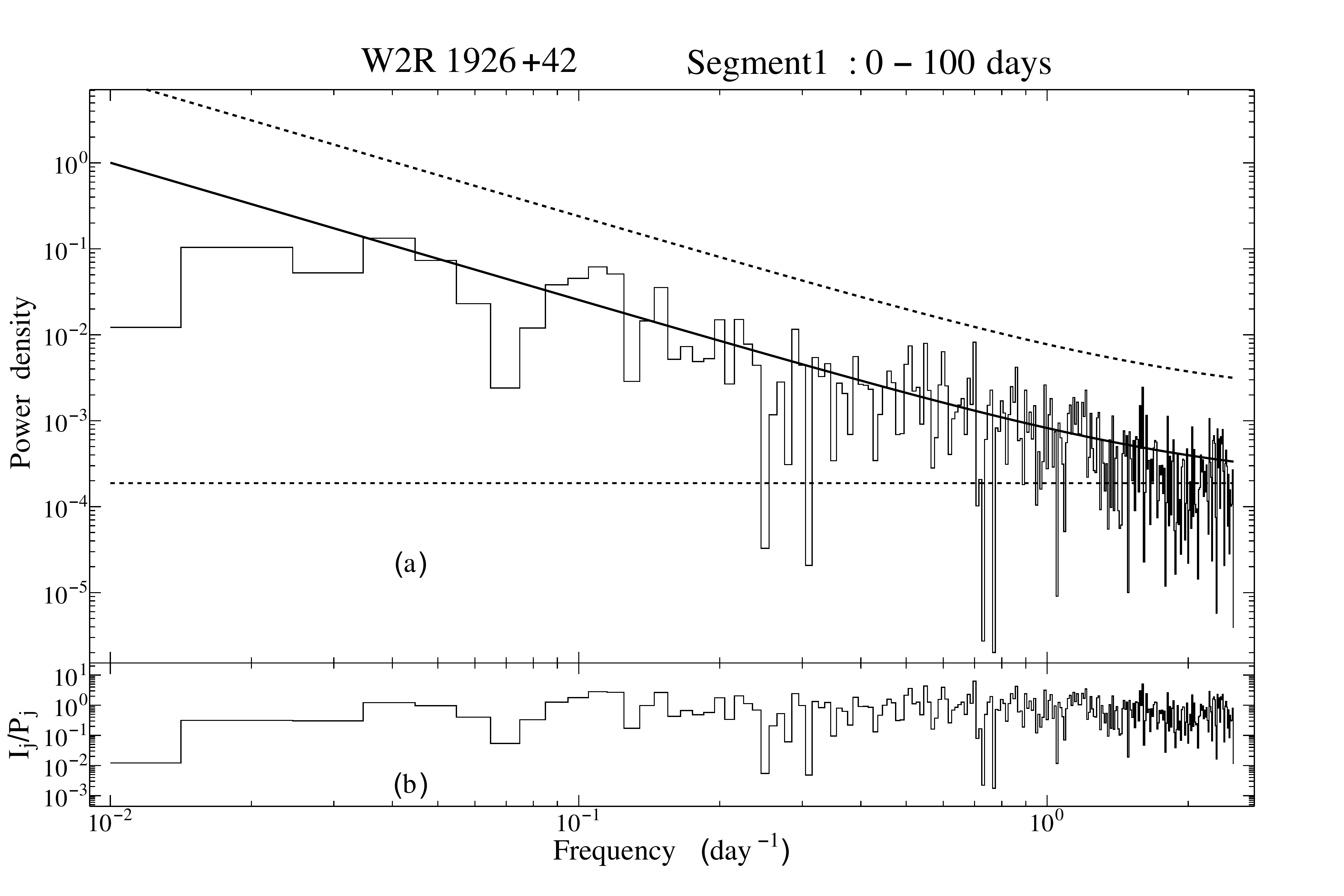}\includegraphics[scale=0.2]{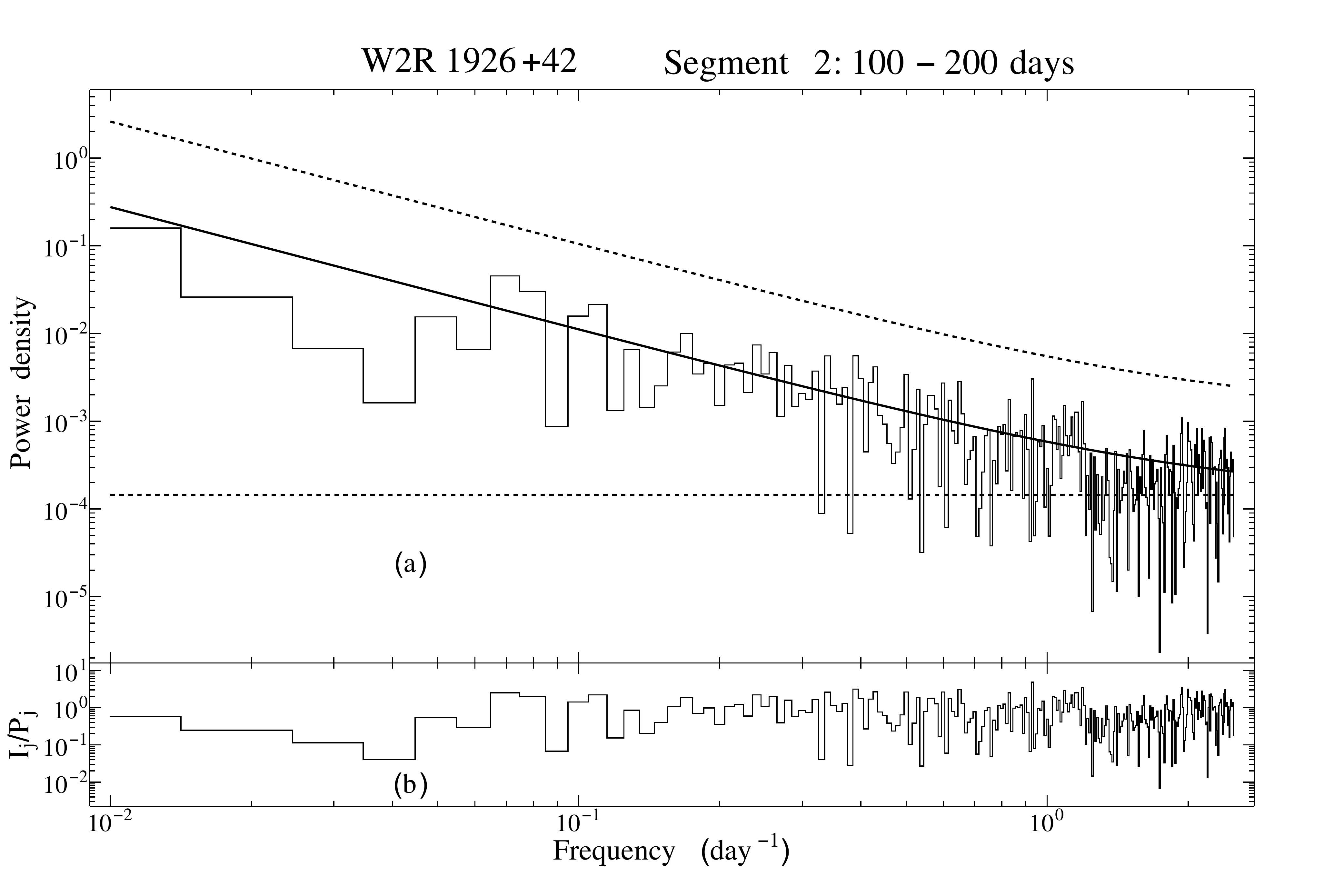}}
\centerline{\includegraphics[scale=0.2]{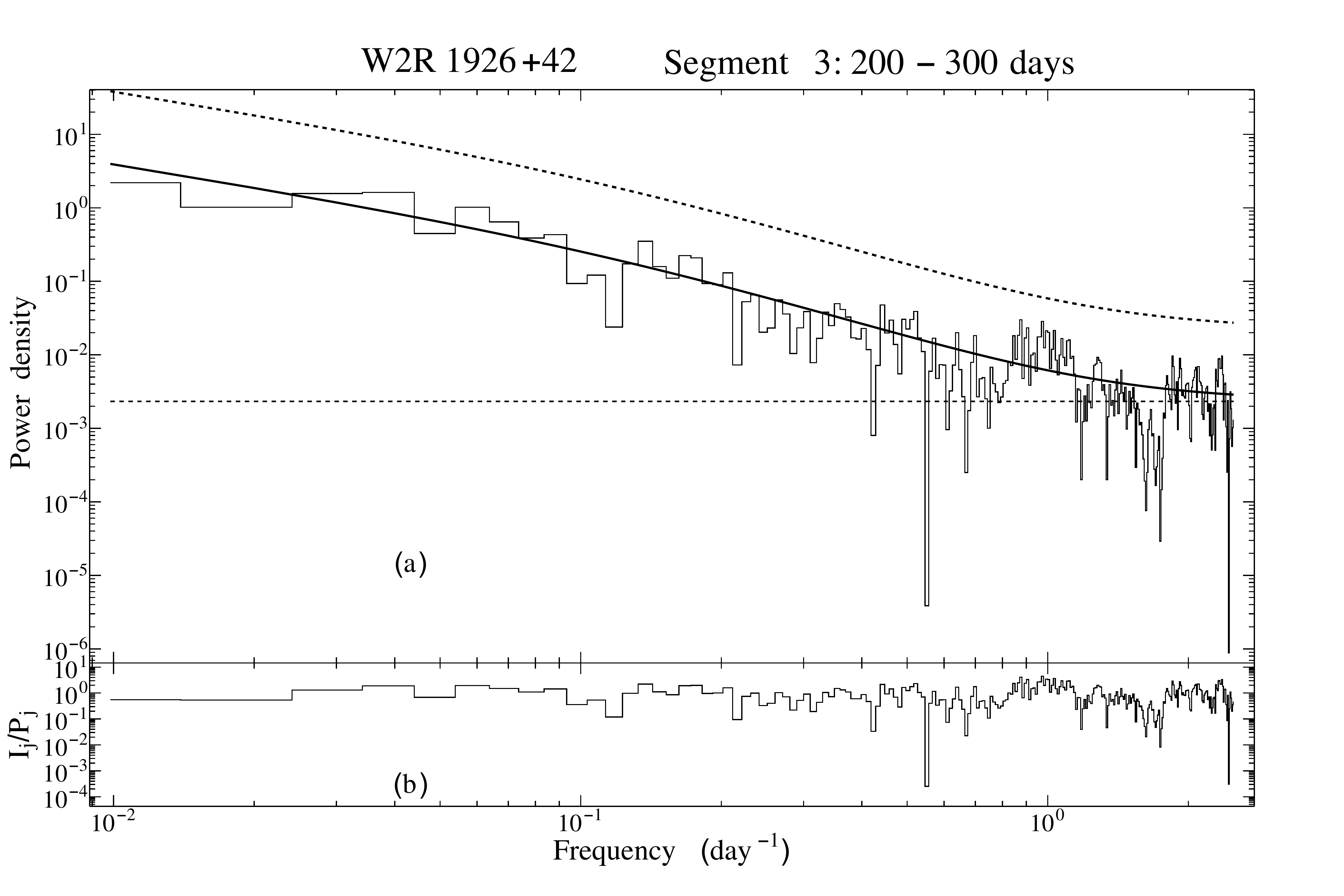}\includegraphics[scale=0.2]{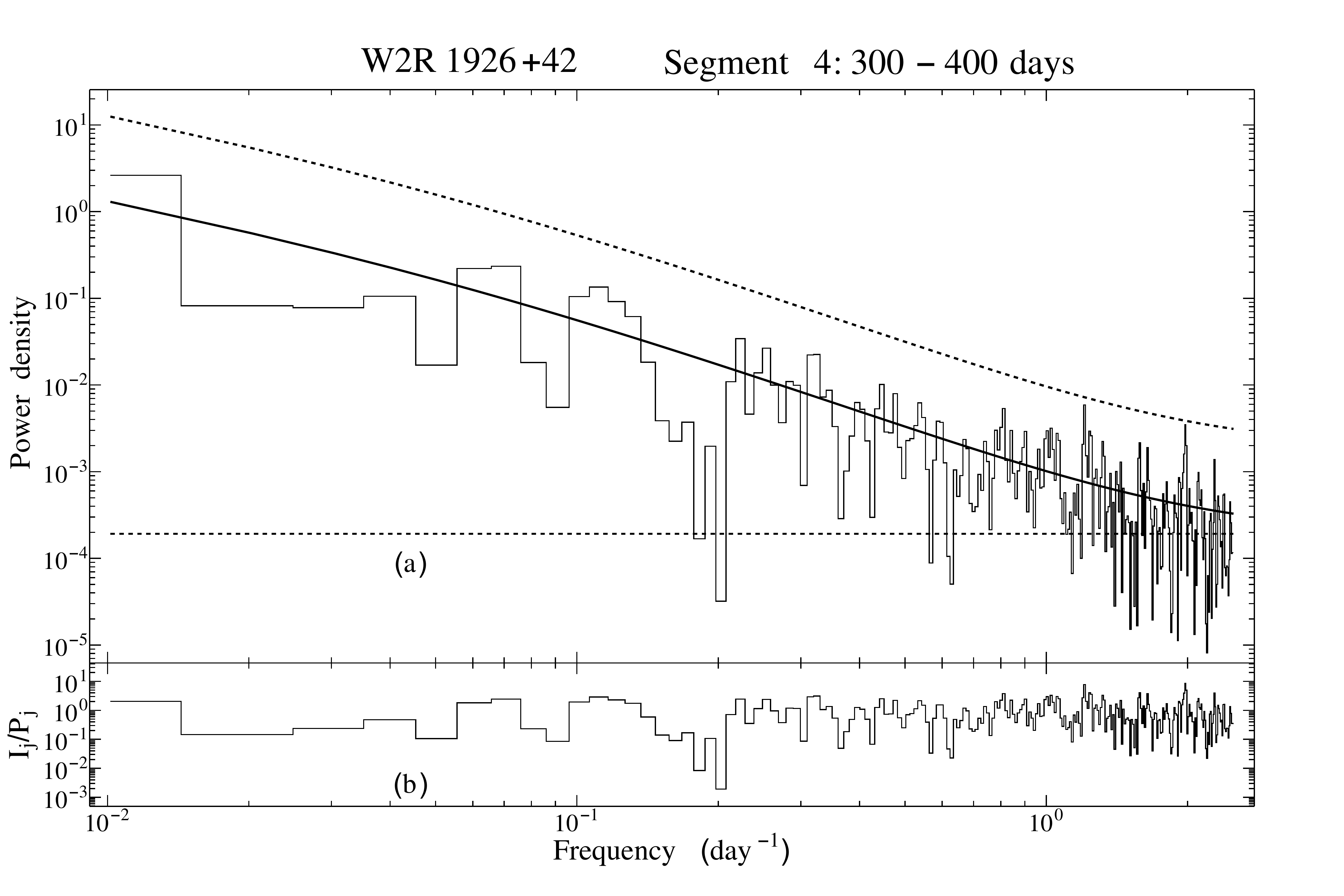}}
\centerline{\includegraphics[scale=0.2]{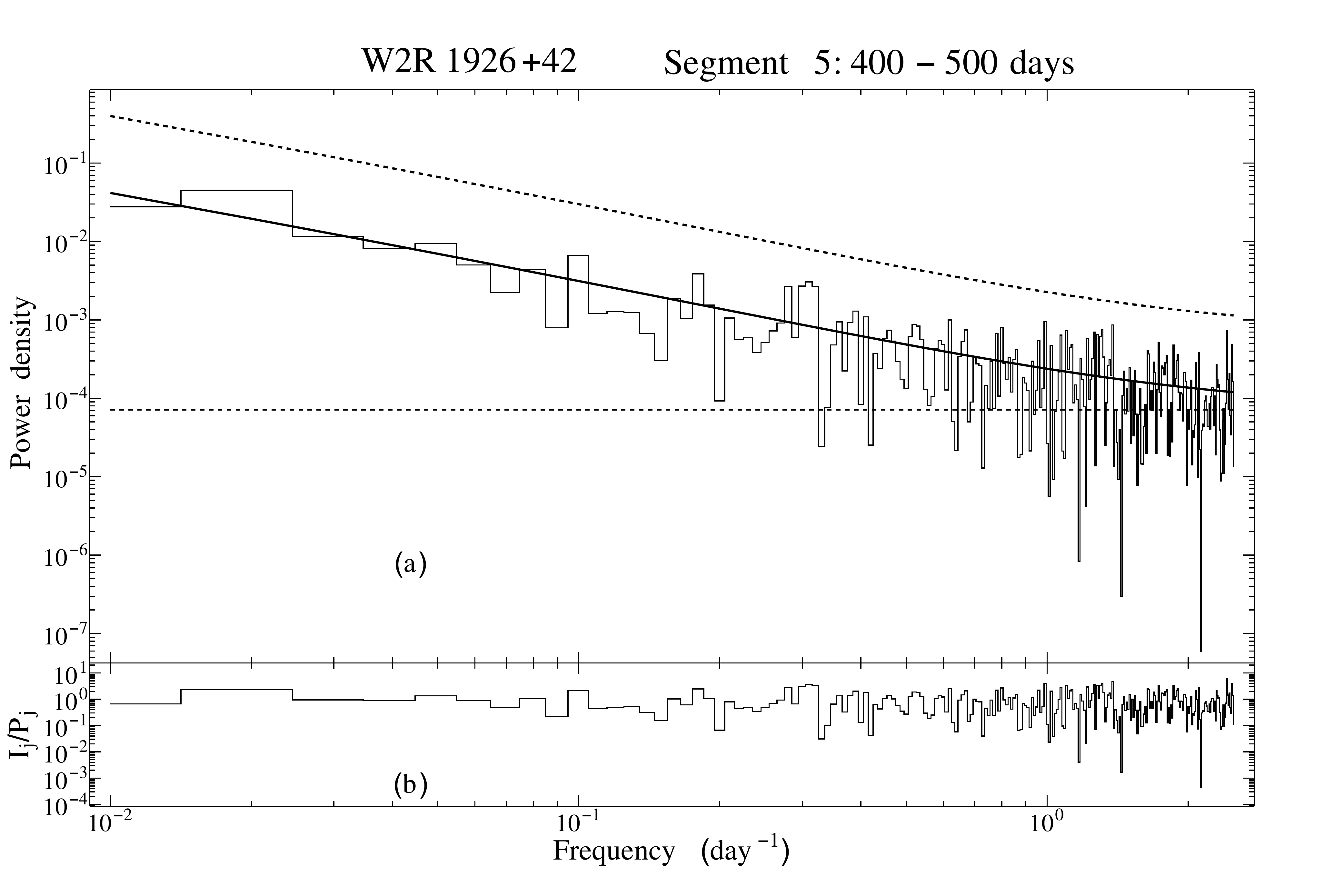}\includegraphics[scale=0.2]{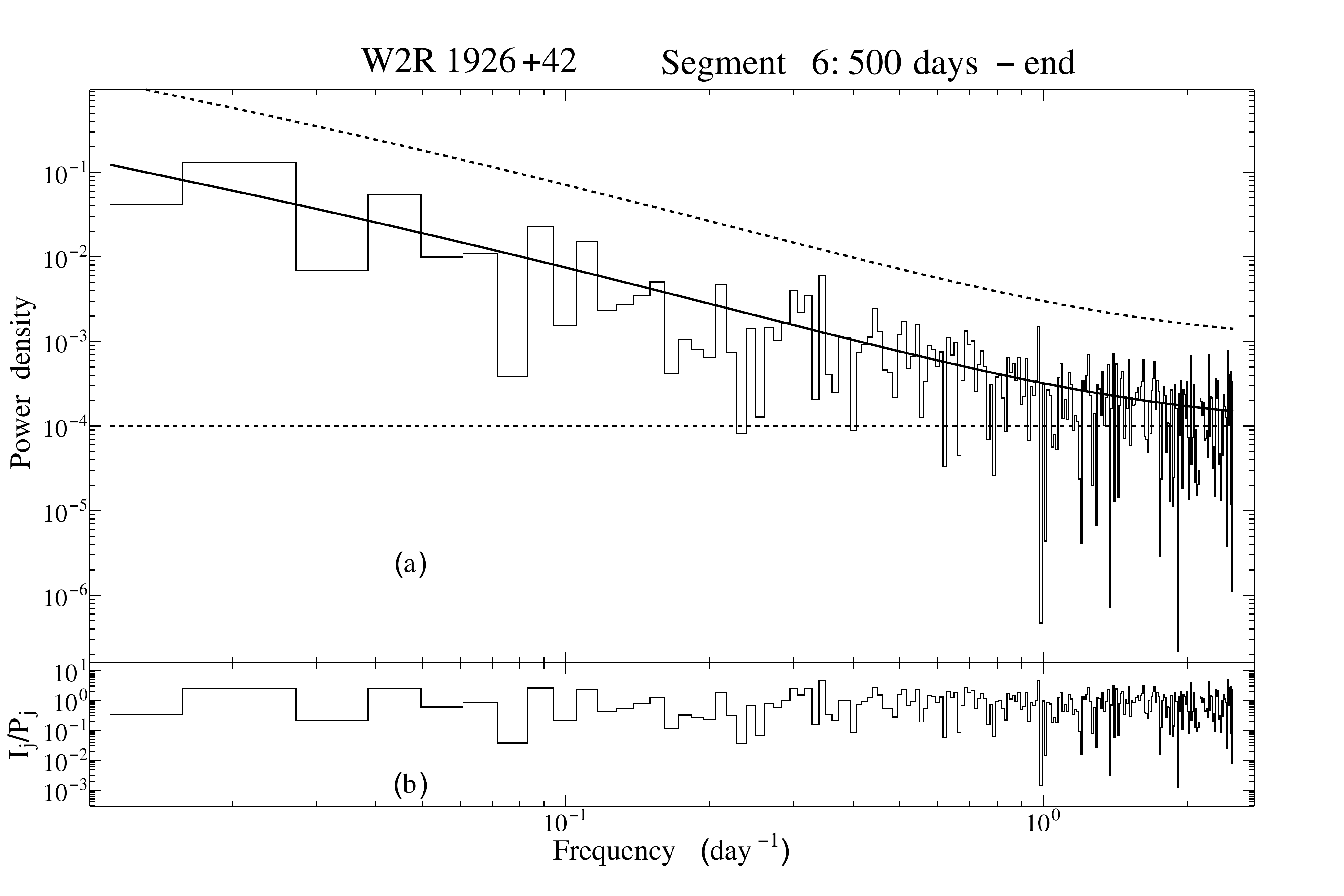}}
\caption{Periodogram analysis of segments 1-6. The best fit model is the solid curve, the dashed curve above it is the 99 \% significance contour which can identify statistically significant quasi-periodic components, the {\bf dashed} horizontal line is the white noise level and the plot below each periodogram panel shows the fit residuals.}
\label{seg16psd}
\end{figure}

The analysis of the full light curve favours the power law shaped PSD shape with slope $-1.4 \pm 0.1$. No statistically significant QPO is detected using the periodogram analysis for any of the individual segments or the entire light curve. Detailed results are presented in Table \ref{PSD}. The periodogram analysis for all segments is presented in Fig. \ref{seg16psd} where the best fit PSD shape is over-plotted on each binned periodogram. The power law model describes the PSD shape well in 2/6 segments (1 and 2), the slope ranging between $-1.2$ and $-1.8$ across all segments with a weighted mean slope of $-1.5 \pm 0.2$, consistent with the inferred slope in radio, optical and X-ray studies (e.g. \citealt{2012MNRAS.425.1357G,2012A&A...544A..80G,2014ApJ...791...74M}), a slope of $-1.8$ reported for this source in \cite{2013ApJ...766...16E} and the range of $-1.2$ to $-2.0$ inferred from a study of four other radio loud AGN in the Kepler field in \cite{2014ApJ...785...60R}. Of the remaining 4 segments, in three segments (3, 4 and 6), both the power law and the bending power law have a comparable probability. These models may thus not be able to adequately describe the PSD shape in these segments. A similar conclusion was reached in the analysis of the combined light curve of the first two segments in \cite{2013ApJ...766...16E} where both PSD models underestimated the power at the highest frequencies. It is thus useful to model the PSD shape using statistical models such as the damped random walk (e.g. \citealt{2009ApJ...698..895K,2010ApJ...708..927K,2010ApJ...721.1014M,2014ApJ...786..143S}) or the CARMA process \citep{2014ApJ...788...33K} where the PSD shape is described effectively by a sum of Lorentzians, which we plan to address in future work. The bending power law is the best fit for the segment 5 light curve where a bend frequency with an associated timescale $T_B = 6.2^{+6.4}_{-3.1}$ hours (ranging between 3.2 - 12.6 hours) is inferred, consistent within errors of a bend timescale of $\sim$ 4 hours inferred for the same source in \cite{2013ApJ...766...16E}. 

\subsection{Wavelet analysis}

\begin{figure}
\centerline{\includegraphics[scale=0.4]{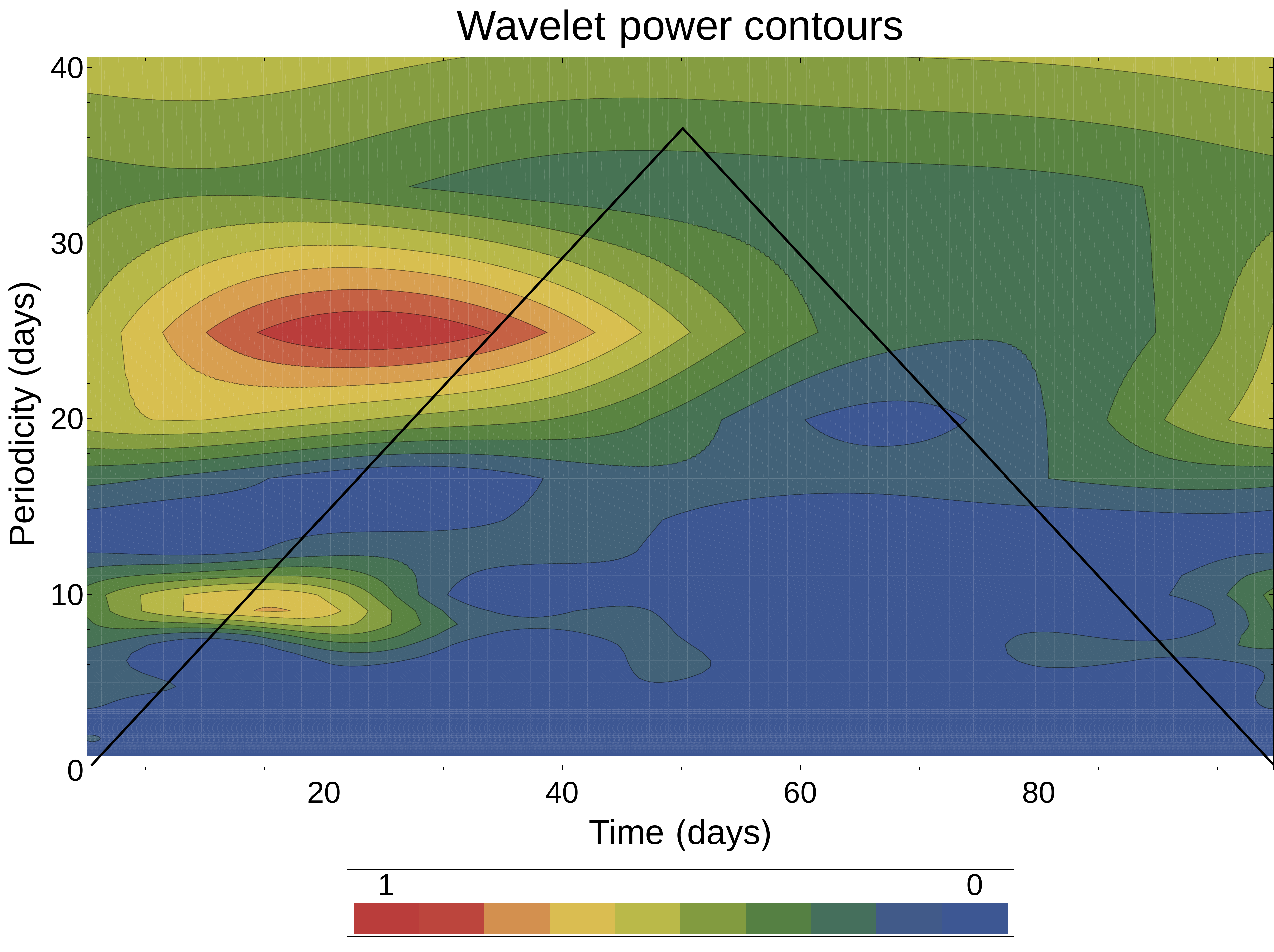}}
\caption{Wavelet analysis of segment 1 light curve (0-100 days) showing strong QPO features peaked at 9.1 days and 24.1 days. Red contours are highly significant levels in the power spectrum and blue contours correspond to the average background power. The power spectrum peaks outside the cone of influence (black triangle shaped region) could be affected by edge effects due to the cyclic nature of the sampling wavelet process.}
\label{waveletseg1}
\end{figure}

The wavelet analysis was conducted for the full light curve and each of the six segments to search for statistically significant QPOs and their properties. Here, we report only those features which indicate a statistical significance of $>$ 90\%. In segment 1 (0 - 100 days), a possible QPO peaked at 9.1 days lasting for the first 30.4 days (3.4 cycles) is inferred with a statistical significance of $>$ 99.9 \%. In the same light curve, another broad feature peaked at 24.9 days is inferred, though with a lower significance of 97.5 \% which lasts for a duration of only 11.2 days (0.5 cycles). This could be a harmonic of the 9.1 days QPO considering that the broad feature is between $\sim$ 6.5 - 12 days. No statistically significant QPO is detected in any of the other individual segments. The analysis of the full light curve indicates a possible QPO peaked at 15.9 days lasting for 84 days (5.3 cycles) with a statistical significance of 93.3 \% and another at 30.9 days lasting for 84 days (2.7 cycles) with a statistical significance of 90.2 \%, the later possibly being a harmonic of the former. The wavelet analysis of the segment 1 light curve is presented in Fig. \ref{waveletseg1}.

\section{Discussion}

An observational relation between $F_{\rm var}$ and the mass of the central supermassive black hole, $M_{\bullet}$ of the form $F_{\rm var} \propto M^{-0.5}_{\bullet}$ is inferred for Seyfert galaxies (e.g. \citealt{2001MNRAS.324..653L,2003MNRAS.343..164B,2004MNRAS.348..207P,2009MNRAS.394.2141N,2014ApJ...791...74M}). There are hints that the relation is also applicable to other AGN such as blazars (e.g. \citealt{2005ApJ...629..686Z}). Disk based line emission in blazars is rarely observed as relativistic beaming masks most of the emission. Though, during a quiescent phase, when jet based synchrotron contribution to the optical/ultraviolet-soft X-rays part of the spectral energy distribution is fully accounted for, there is still a residual excess emission which can be ascribed to disk based processes (e.g. \citealt{2007A&A...473..819R,2009A&A...507..769R}). The absence of optical polarization during the quiescent phase is also likely from disk based thermal emission (e.g. \citealt{2011ApJ...735...60P}). If the optical emission during the quiescent phase contains a portion of this disk emission, the $F_{\rm var}$ measured here indicates upper limits on the variability due to disk contribution. If we assume that variability in this source during the quiescent phases could be contributed to by disk based processes, we can then apply these empirical relations to infer the black hole mass using the $F_{\rm var}$ ranging between 1.8 \% and 6.8 \%. The black hole mass estimated using $F_{\rm var}$ then gives upper limits.

A relation $\log(F^2_{\rm var}) = (5.08 \pm 0.11) - \log(M_0)$ is suggested by \cite{2003MNRAS.343..164B} where $M_0 = M_{\bullet}/M_{\odot}$. Using this and the above $F_{\rm var}$ range gives $M_{\bullet} = (1.7 - 22.8) \times 10^7 M_{\odot}$. A relation $\log(F^2_{\rm var}) = -2.09 - 1.03 \log(M_7)$ is suggested by \cite{2012A&A...542A..83P} where $M_7 = M_{\bullet}/10^7 M_{\odot}$ for a sample consisting of 161 radio-quiet AGN. Using this and the same $F_{\rm var}$ range gives $M_{\bullet} = (2.0 - 47.8) \times 10^7 M_{\odot}$.

The Doppler factor $\delta$ due to the relativistic time dilation effect ranges between $1.1 - 24.0$, obtained from long term studies of BL Lacertae objects in radio and optical wavelengths (e.g. \citealt{2009PASJ...61..639F,2009A&A...494..527H}) with the distribution peaking at $\sim$ 5. Using this typical $\delta = 5$, the bend timescale in the source frame $\delta T_B/(1+z)$ corrected for the cosmological redshift $z$ ranges between $13.4 - 52.4$ hours ($\sim$ 0.6 - 2.2 days). A study of radio quiet AGN indicates a relation between the bend timescale, the black hole mass $M_{\bullet}$ and the accretion rate (normalized to the Eddington rate) $\dot{m}$, $T_B \propto M_\bullet/\dot{m}$ \citep{2006Natur.444..730M}. Assuming that this relation holds good for W2R 1926+42 and using the Doppler corrected $\delta T_B$ in the relation $T_B \sim 3.33 \times 10^3 M_7/\dot{m}$ \citep{2015arXiv150607769P} and $\dot{m} = 0.1 - 0.3$ \citep{2009MNRAS.399.2041G}, we obtain $M_{\bullet} = (1.5 - 5.9) \times 10^7 M_\odot$. The range obtained is tighter, with the upper limit being one order of magnitude lower than the upper limit obtained using the $F_{\rm var} - M_\bullet$ relation. Though, as argued earlier, since this bend timescale arises in the segment 5 where the source is in a quiescent phase, its origin could be disk related and hence, the above obtained tighter range of black hole mass is more relevant as low bend timescales generally arise due to disk based processes near the innermost stable circular orbit where inflowing plasma makes a transition from the disk edge into the black hole \citep{2014ApJ...791...74M}. The estimated mass range of $(1.5 - 5.9) \times 10^7 M_\odot$ is within the range of estimates made or compiled by previous studies of radio loud AGN ($10^6 - 10^{10} M_\odot$, e.g. \citealt{2008ApJ...685..801Y,2009ApJ...698..895K,2010MNRAS.405..387G,2012MNRAS.424..393F,2012NewA...17....8G,2013ApJ...779..187K}). 

Taking the same typical $\delta$ value, the mean outburst lifetime in the source frame $\delta \tau/(1+z)$, ranges between $17.3 - 100.4$ days. Using this, we can estimate an upper limit on the region of origin of the variability, at a distance $r$ from the central black hole, given by the relation $r \leq c \delta \tau/(1+z)$. This gives $r \leq 2.6 \times 10^{15} ~{\rm m} ~\leq 1.75$ pc which indicates that the outburst region is closer to the jet launching region than that indicated through radio observations of blazars which typically indicate $r_{\rm core} >$ a few to tens of pc (e.g. \citealt{2015MNRAS.452.2004M}). The shock in jet model (e.g.  \citealt{1985ApJ...298..114M}) proposes that emitting cores along the jet tend to flare when a relativistic shock propagates down the jet and causes the core to brighten. In this model, the distance from the jet base, $r \propto \nu^{-1}$ \cite[e.g. ][]{1981ApJ...243..700K} where $\nu$ is the observation frequency. Our result of $r \leq 1.75$ pc is then qualitatively consistent with this model which was inferred to explain multi-wavelength radio variability (4.8 GHz - 36.8 GHz) based $r \sim 1.98 - 69.21$ pc \citep{2015MNRAS.452.2004M}. Thus, the origin of the flare could be from a shock interaction in the jet.

Both disk and jet based processes can lead to quasi-periodic variability. Disk oscillations could be due to radial perturbations to the plasma inflow, acoustic oscillations (e.g. \citealt{1997ApJ...476..589P}) which could cause variability over a timescale of the order of the dynamic timescale $T_V \sim 2 \pi G M_\bullet r^{3/2}/c^3$, a few thousand to ten thousand seconds, which corrected for $\delta$ is $\leq$ $1/(1+z)$ days, less than that inferred from this study. Orbital features on the disk inflowing into the black hole can also cause variability (e.g \citealt{1993ApJ...406..420M}), though over similar timescales as the dynamic timescale, thus, unlikely to explain the observed variability. Other disk based processes include the quasi-periodic injection of plasma into the jet base from the disk which could be interpreted as a disk-jet connection and could be caused by disk oscillations, excited by a binary black hole \citep{2006ApJ...650..749L,2013MNRAS.434.3487A,2014MNRAS.443...58W} and are accompanied by the detection of multiple harmonics. Though, the timescales involved are of the order of a few years and are hence unlikely to be applicable here. 

Jet based variability processes include shocks in the jet \citep{1979ApJ...232...34B}, turbulent relativistic flow in the jet accompanying shocks \citep{2014ApJ...780...87M}, jet precession due to the Lense-Thirring effect or vertical oscillations set up by a warped disk formed in a binary black hole scenario (e.g. \citealt{1992A&A...255...59C,2004ApJ...615L...5R,2013ApJ...765L...7N}), preferential beaming of the shock front \citep{1992A&A...259..109G} or relativistic orbital features in the jet \citep{2014JApA...35..431M,2015ApJ...805...91M}. Typical timescales obtained in these models are of the order of a few days to years. In the shock and turbulence based processes, the timescales are not coherent and are generally aperiodic. If we assume that the QPO at 9.1 days is a true feature, the Doppler factor and cosmological redshift factor corrected timescale $\delta T/(1+z)$ is 39.4 days. If the origin of the QPO is due to emission from orbital features in the jet (e.g. \citealt{1992A&A...255...59C,2004ApJ...615L...5R,2015ApJ...805...91M}), the distance from the central engine can be calculated using the Keplerian relation $r = (\delta T c^3/(2 \pi G M_\bullet (1+z)))^{2/3}$, which gives $r = (152 - 378) ~ G M_\bullet/c^2$ for the above $T$ and $M_\bullet = (1.5 - 5.9) \times 10^7 M_\odot$. In this scenario, the relativistic orbiting features can be launched from a region $\sim 150 G M_\bullet/c^2$ away from the black hole and can make a few orbits along a helical jet as it beams emission towards the observer line of sight, thus causing the QPO \citep{2015ApJ...805...91M}. In that study, typical timescales of a few days and a PSD slope $\sim -2$ are obtained from simulated light curves from a jet based orbital process. Further, it is seen that there are regimes of jet parameter choices where the QPO is weak and lasts only for a few cycles or is completely absent for emission from a region consisting of multiple orbiting features. In this study, as the QPO is weak and exists only for a short duration and the typical PSD slope of $-1.5 \pm 0.2$ is consistent with the simulations of \cite{2015ApJ...805...91M}, our results are likely to be due to variability from orbital features in the jet. As there are outburst and quiescent phases following the segment 1 light curve without any coherent features, the observed variability over the entire duration could be caused by both orbital features and other jet based aperiodic processes such as shock in jet and turbulence. 

\section{Summary and conclusion}

Our analysis of the long term Kepler light curve of W2R 1926+42 ($\sim$ 1.6 years) indicates a variety of variability properties during the different intervals of observation. Below, we summarize our study.

\begin{enumerate}
\item The normalized excess variance, $F_{\rm var}$ is studied as a function of observation interval. It is found to follow the same pattern as that of the variability in the light curve indicating a relation between the variance and the flux. Our study of the $F_{\rm var} -$ flux relation using a longer light curve as compared to \cite{2013ApJ...766...16E} helps answering whether the deviation in the linear relation inferred in their study is a true feature. We find the deviation previously reported to be a fluctuation from the mean. The inferred relation in our study is consistent with linearity. 
\item The $F_{\rm var}$ ranges between 1.8 \% in the quiescent phase to 43.3 \% during the outburst. Applying an empirical relation $F_{\rm var} \propto M^{-0.5}_\bullet$ to the $F_{\rm var}$ in the quiescent phase, we obtain a mass $M_\bullet = (1.7 - 47.8) \times 10^7 M_\odot$.
\item From the mean outburst lifetime in the source frame of $17.3 - 100.4$ days, the region from which the outburst occurs is at $r \leq 2.6 \times 10^{15} ~{\rm m} ~\leq 1.75$ pc indicating that it is close to the jet launching region, and is consistent with its origin from a shock in the jet.
\item Time series analysis is conducted using the Fourier power spectrum and the wavelet analysis methods to study the PSD shape and to infer statistically significant QPOs.
\item The weighted mean slope of the best fitting power law PSD model is $-1.5 \pm 0.2$ which lies within the range of previously inferred slopes for this and other sources.
\item The bending power law is the best fit for the segment 5 light curve with an associated timescale of $T_B = 6.2^{+6.4}_{-3.1}$ hours (3.2 - 12.6 hours). Using a relation $T_B \propto M_\bullet/\dot{m}$, with $T_B$ corrected for the cosmological redshift $z$ and the Doppler factor $\delta$, we obtain a tighter limit on mass $M_\bullet = (1.5 - 5.9) \times 10^7 M_\odot$. This is taken as the representative range as the emission and variability following the flare in the quiescent phase can be due to disk based processes. This is the first mass measurement for this source using multiple empirical relations.
\item A possible QPO peaked at 9.1 days and lasting 3.4 cycles is inferred from the wavelet analysis of the segment 1 light curve. The origin of the variability is from a region $(152 - 378)~ G M_\bullet/c^2$ away from the jet base near the central black hole. 
\end{enumerate}

Assuming that the QPO is a true feature and supported by the other timing analysis products such as the PSD slope, the variability in segment 1 and in general could be due to jet based processes including orbital features in a relativistic helical jet (preferentially beamed emission for a few cycles causing the weak and short duration QPO) and others such as shocks and turbulence. The sequence of inferences with observation time is thus the identification of a possible QPO peaked at 9.1 days lasting 3.4 cycles in the initial quiescent phase using which $r = (152 - 378)~ G M_\bullet/c^2$; the outburst with a mean lifetime of 13.6 $\pm$ 9.6 days using which $r\leq 1.75$pc suggesting a shock based origin of the outburst; a quiescent phase, the $F_{\rm var}$ range during which is used to determine $M_\bullet$ ranges and during which a possible disk processes based bend timescale of $6.2^{+6.4}_{-3.1}$ hours is used to determine a tight range of $M_\bullet = (1.5 - 5.9) \times 10^7 M_\odot$.

\section*{Acknowledgements}
We thank the anonymous referee for useful comments which helped us clarify our analysis methodology and helped improve the presentation of our work. This research was partially supported by Scientific Research Fund of the Bulgarian Ministry of Education and Sciences under grants DO 02-137 (BIn 13/09) and NTS BIn 01/9 (2013). This paper includes data collected by the Kepler mission. Funding for the Kepler mission is provided by the NASA Science Mission directorate. The authors would like to thank the Kepler mission team for the use of this data set in our analysis.

\bibliography{bibliography}
\onecolumn

\begin{table}
\begin{tabular}{cccc}\hline
Duration & Phase & $F_{\rm var}$ & Characteristic \\
 (days)  & Name  & range (\%)    & Timescale/s \\
         &       &               & (days)      \\ \hline
0 - 180  & Quiescent? & 4.8 - 16.6 & - \\
 & & & \\
180 - 250 & Pre-outburst & 9.9 - 22.7 &  $\tau_R = 27.6 \pm 7.4$ \\
 & & & (Outburst rise time) \\
250 & Outburst & 43.3 & - \\
 & & & \\
250 - 380 & Post-outburst& 7.7 - 17.8 & $\tau_D = 7.0 \pm 5.1$ \\
 & & & (Outburst decay time)\\
380 - 589 & Quiescent & 1.8 - 6.8 & - \\ 
 & & & \\
0 -589 & - & 1.8 - 43.3 & $\tau = 13.6 \pm 9.6$ \\
 & & & (Outburst lifetime) \\
\hline
\end{tabular}
\caption{$F_{\rm var}$ range in segments of Fig. \ref{Fvart} and inferred timescales.}
\label{Fvar}
\end{table}     

\begin{table}
\centering
\scalebox{.9}{
\begin{tabular}{ccccccccccc}
\hline
Segment &           & \multicolumn{5}{c}{Periodogram analysis}                & \multicolumn{4}{c}{Wavelet analysis} \\
No.     & PSD       & \multicolumn{3}{c}{PSD Fit parameters}    & AIC & Model & Periodicity & Statistical & Duration & No. of  \\
        & model     & \multicolumn{3}{c}{}                      &     & likelihood & (days) & significance& (days)   & cycles  \\
        &           & log(A) & $\alpha$ & log(f$_b$)            &     &            &        &             &          &        \\ \hline \\
1. & {\bf PL} & -3.2 $\pm$ 0.1 & -1.6 $\pm$ 0.1 & & 2958.52 & 1.00 & 9.1 & $>$99.9 & 30.4 & 3.4 \\
(0-100 days) & BPL & -2.6 $\pm$ 0.1 & -2.6 $\pm$ 0.5 & -0.26 $\pm$ 0.22 & 2967.33 & 0.02 & 24.9 & 97.5 & 11.2 & 0.5 \\  \\
2. & {\bf PL} & -3.4 $\pm$ 0.1 & -1.4 $\pm$ 0.2 & & 3166.12 & 1.00 & 14.2 & 92.7 & 8.4 & 0.6 \\
(100-200 days) & BPL & -3.0 $\pm$ 0.1 & -3.0 $\pm$ 0.5 & -0.26 $\pm$ 0.25 & 3171.33 & 0.07 \\ \\
3. & PL & -2.4 $\pm$ 0.1 & -1.7 $\pm$ 0.1 & & 1916.51 & 0.77 & - & - & - & - \\
(200-300 days) & BPL & -1.4$\pm$ 0.2 & -2.2 $\pm$ 0.4 & -0.82 $\pm$ 0.30 & 1915.97 & 1.00 \\ \\
4. & PL & -3.1 $\pm$ 0.1 & -1.8 $\pm$ 0.1 & & 2773.56 & 0.94 & - & - & - & - \\
(300-400 days) & BPL & -1.8 $\pm$ 0.2 & -2.0 $\pm$ 0.2 & -1.25 $\pm$ 0.30 & 2773.44 & 1.00 \\ \\
5. & PL & -3.7 $\pm$ 0.1 & -1.2 $\pm$ 0.1 & & 3591.81 & 0.34 & - & - & - & - \\
(400-500 days) & {\bf BPL} & -3.3 $\pm$ 0.1 & -1.5 $\pm$ 0.6 & -0.59 $\pm$ 0.31 & 3589.66 & 1.00 \\ \\
6. & PL & -3.7 $\pm$ 0.1 & -1.6 $\pm$ 0.1 & & 3051.00 & 0.83 & - & - & - & - \\
(500-600 days) & BPL & -2.8 $\pm$ 0.2 & -1.7 $\pm$ 0.5 & -1.21 $\pm$ 0.3 & 3050.62 & 1.00 \\ \\
1-6   & {\bf PL} & -2.6 $\pm$ 0.1 & -1.4 $\pm$ 0.1 & & 1399.56 & 1.00 & 15.9 & 93.3 & 84.0 & 5.3 \\
(All) & BPL & -2.0 $\pm$ 0.1 & -3.5 $\pm$ 0.4 & -0.95 $\pm$ 0.14 & 1413.82 & $10^{-3}$ & 30.9 & 90.2 & 84.0 & 2.7 \\ \hline
\end{tabular}}
\caption{Results from the parametric PSD models fit to the periodogram. Columns 1 -- 11 give the segment number, the model (PL: power law + constant noise, BPL: bending power law + constant noise), the best-fit parameters $\log(N)$, slope $\alpha$ and the bend frequency $f_b$ with their 95\% errors derived from $\Delta S$, the Akaike information criteria value (AIC), likelihood of a particular model, inferred QPO timescale from the wavelet analysis (in days), its statistical significance based on a Monte-Carlo simulations based test, its duration of existence (in days) and the number of cycles it lasts for. The best fit PSD model is highlighted in bold face.}
\label{PSD}
\end{table}

% Don't change these lines
\bsp	% typesetting comment
\label{lastpage}
\end{document}